\documentclass[apj, iop]{emulateapj}

 \usepackage{amsmath}
 \usepackage{url}

\shorttitle{Fast electron beam induced intense radio emission}
\shortauthors{Yu et al.}

\begin{document}

\title{Electron beam induced radio emission from ultracool dwarfs}

\author {S. Yu, J.G. Doyle}
\affil{Armagh Observatory, College Hill, Armagh BT61 9DG, N. Ireland}
\author { A. Kuznetsov}
\affil{Armagh Observatory, College Hill, Armagh BT61 9DG, N. Ireland}
\affil{Institute of Solar-Terrestrial Physics, Irkutsk 664033, Russia}
\author {G. Hallinan}
\affil{National Radio Astronomy Observatory, 520 Edgemont Road, Charlottesville, VA 22903, USA}
\affil{Department of Astronomy, University of California, Berkeley, CA 94720, USA}
\author {A. Antonova}
\affil{Department of Astronomy, St. Kliment Ohridski University of Sofia, 5 James Bourchier Blvd., 1164 Sofia, Bulgaria}
\author {A.L. MacKinnon}
\affil{School of Physics and Astronomy, University of Glasgow, Glasgow G12 8QQ }
\author {A. Golden}
\affil{Price Center, Albert Einstein College of Medicine, Yeshiva University, Bronx, NY 10461}


\begin{abstract}
We present the numerical simulations for an electron-beam-driven and loss-cone-driven 
electron-cyclotron maser (ECM) with different plasma parameters and different magnetic 
field strengths for a relatively small region and short time-scale in an attempt to 
interpret the recent discovered intense radio emission from ultracool dwarfs. We find that a 
large amount of electromagnetic field energy can be effectively released from the 
beam-driven ECM, which rapidly heats the surrounding plasma. A rapidly developed high-energy 
tail of electrons in velocity space (resulting from the heating process of the ECM) may produce
the radio continuum depending on the initial strength of the external magnetic field and 
the electron beam current. Both significant linear polarization and circular polarization of 
electromagnetic waves can be obtained from the simulations. The spectral energy distributions 
of the simulated radio waves show that harmonics may appear from 10 to 70$\nu_{\rm pe}$ ($\nu_{\rm pe}$ 
is the electron plasma frequency) in the non-relativistic case and from 10 to 600$\nu_{\rm pe}$ 
in the relativistic case, which makes it difficult to find the fundamental cyclotron 
frequency in the observed radio frequencies. A wide frequency band should therefore be covered by 
future radio observations. 
\end{abstract}


\section{Introduction}
\label{intro}

Ultracool dwarfs (UCDs) are those objects with spectral 
type later than M7 and low luminosity. Recent observations of 193 UCDs reveal that 
12 UCDs produce intense radio emission with flux 
densities up to hundreds of $\mu$Jy (e.g. 
\citet{Berger01,Berger02,Berger05,Burgasser05,Hallinan06,Antonova08,
Mclean11b}). Some of them show a highly circularly 
polarized radio pulse with regular periods, and a flux density 
up to 15 mJy \citep{Hallinan07,Berger09}. 

In a range of past studies, these radio features of UCDs have been presented as 
a function of magnetic field, spectral type, rotation, age, binarity, and 
association with the X-ray and H$\alpha$ emission (e.g. \citet{Berger05,
Hallinan06,Antonova08,Berger10,Mclean11b}). The regular periods of radio 
pulses from TVLM 513-46546 and the L dwarf binary 2MASSW J0746425$+$200032 
indicate that the radio activity of UCDs is strongly in conjunction with 
their rotation \citep{Hallinan07,Berger09} which is one of the crucial 
factors to influence the magnetic field by a differential rotation induced 
dynamo theory \citep{Parker55} although other mechanisms, such as a 
turbulence-induced dynamo \citep{Durney93} (for small-scale fields) or 
$\alpha^{2}$ dynamo \citep{Chabrier06} (for large-scale fields), can also 
contribute to the magnetic field. 
The topology of the magnetic field on UCDs may be understood as a 
dipole due to the narrow bunching of multiple pulses of both left- and 
right- 100\% polarization \citep{Hallinan07}. \citet{Berger09}, however, has suggested that
the field topology maybe more complex - due to a 1/4 phase lag of the radio pulses 
compared to H$\alpha$. This is based on the assumption that the emission is 
parallel to the magnetic field, but if the emission is perpendicular to the field, then the 1/4 
phase lag is in agreement with a dipole magnetic field geometry. Hence a determination 
of the magnetic field and its structure is critically important. 

Precise analysis of the time domain of radio emission from TVLM 513-46546 \citep{Doyle10} 
and sporadic radio emission from UCDs \citep{Antonova07} show that 
large-scale fields may be stable on UCDs for long periods, from a few months 
to years. The steady magnetic fields on UCDs are also confirmed by the multi-frequency observations 
of a late-M dwarf binary \citep{Osten09}. The field strength can be determined using two specific radiation mechanisms  
- gyrosynchrotron or electron cyclotron maser (ECM). The first mechanism suggests 
a field strength in the range of 0.1$-$1000 G \citep{Berger02,Berger06}, 
while the latter implies a kG field. However, the form of the frequency-field strength 
relation becomes complicated when the ECM mechanism is applied to a many electron system since (1) 
the absorption and emission of different layers in the magnetosphere (or atmosphere) 
would be significant due to the different plasma environments and magnetic field 
configuration (e.g. see the discussion of the gyromagnetic absorption in \citet{Melrose82}); and  
(2) as we will show in this paper, for the motion of a group of electrons in a magnetic 
field, the ECM can generate a multiple peak structure for the spectral energy distribution. 
Hence, coverage of the full dynamic radio spectrum, including the low frequency band (hundreds 
of MHz) and the very high frequency band, are important for a proper understanding of the 
radiation process. 

The observed power-law radio continuum and low level circular polarization 
($<$40\%) from several UCDs, such as for the M8.5 dwarf DENIS 1048-3956 between 
3-30 GHz in four 2 GHz bandwidths \citep{Ravi11}, may be interpreted as 
gyrosynchrotron radiation if the surrounding 
plasmas is optically thin. On the other hand, the high 
brightness temperature ($\sim10^{15}$ K) and highly (up to 100\%) circular polarization 
of the radio pulses \citep{Hallinan07,Berger09} suggest that the dominant emission mechanism 
is the ECM. This mechanism was initially assumed to be driven by a loss-cone velocity 
distribution (\citet{Melrose82} and references therein), but was subsequently developed 
to ring shell distribution or horseshoe distribution \citep{Pritchett85}. 

The operation of the ECM is rather simple, i.e. electrons with an anisotropic 
distribution transversely move in an external magnetic field. This leads to the 
application of the ECM to the radio emission from the solar planets, magnetic-chemically peculiar 
stars (e.g. \citet{Lo12}), to some compact extragalactic radio sources (e.g. \citet{Melrose82,Dulk85,Treumann06} 
and references therein). The generation 
of the auroral kilometric radiation on the Earth has been interpreted in terms of the  
ECM, where the velocity distribution of electrons may not be a loss-cone caused by the 
magnetic mirror effect; but due to a horseshoe distribution associated with the 
acceleration of particles in a magnetic field-aligned electric field \citep{Wu79,Chiu78,Ergun00}. 
A similar interpretation can be applied to the decametric radiation on Jupiter, Saturnian 
kilometric radiation \citep{Zarka98,Zarka04} and solar millisecond microwave spikes 
\citep{Aschwanden90b,Fleishman03}. The ECM can also be a strong candidate for the possible presence of 
radio emission in exoplanets \citep{Zarka07,Griebmeier07,Jardine08}. Furthermore, it can  
be an effective mechanism for the radio-frequency heating of X-ray emitting plasma in 
solar flares \citep{Melrose84b}. Recently, it was suggested that the ECM generated 
by the low-density relativistic plasmas in many fine localized regions can interpret the high 
brightness temperature detected from Blazar jets \citep{Begelman05}. More discussion and application 
of the ECM can be seen in a review in \citet{Treumann06}.  

In fact, gyrosynchrotron radiation and ECM belong to the same family - the motion of 
electrons in a magnetic field. ECM may efficiently heat the surrounding electrons to 
form a high-energy tail or even a bump in the velocity space 
that induces gyrosynchrotron radiation to contribute to the radio continuum. 
Combining the short time-scale and self-quenching features of the ECM, we can also 
understand the high brightness temperature of the radio pulses. Electron beams would 
be common in the context of the astrophysical process since there are plenty of sources 
for generating them. Recent cool atmospheric models indicate that collisions of significant 
volume of molecular clouds may trigger a tempestuous discharge process such as lightning, 
resulting in a high-degree ionization in the local molecules or atoms, and the release of 
a large number of electrons \citep{Helling11}, which increases the probability of magnetic 
reconnection events. The electrons may be released and accelerated from the magnetic 
reconnection or outflow jets indicated by oxygen forbidden emission 
lines \citep{Whelan07}, which might result from the intense activity below the chromosphere 
of UCDs. 

In order to understand the radio emission from ultracool dwarfs and infer the 
magnetic field and the plasma environment, an investigation of a many electron 
system moving in an external magnetic field is essential. Numerical simulations can provide 
the opportunity to obtain the detailed process self-consistently and an interpretation 
for the radio emission. In this paper, we attempt to interpret the radio pulses from UCDs using an 
electron-beam (or current-beam) driven ECM, with concentration on the 
microscopic energy transformation by treating the electron 
population as charged particles in a simulation box. We also investigate 
the growth rate and polarization of the released electromagnetic (EM) waves 
and the spectral energy distribution (SED). In \S\,\ref{sec_simulation}, we 
briefly describe the physical model, numerical method, and the initial 
conditions to carry out the simulations. In \S\,\ref{sec_resultsI} and 
\S\,\ref{sec_resultsII}, we present the results for the non-relativistic beam-driven and 
loss-cone-driven ECM. In \S\,\ref{sec_relativistic}, we present the results for the relativistic beam-driven 
instability. We make a brief comparison with the observations in \S\,\ref{sec_comparison},  
then summarize the simulations and draw conclusions in \S\,\ref{sec_conclusion}.

\section{Configuration of Simulation}
\label{sec_simulation}

\subsection{Physical model}
\label{sec_physics}
 
We assume that electron beams are generated by some intense events 
on UCDs, e.g. magnetic reconnection or jet events. 
When the electron beams move to the magnetosphere of the UCDs, they interact 
with the magnetic field and the surrounding plasmas. In the present study, we 
neglect the influence of heavy ions. Here, we investigate the 
energy transfer, including the induced EM field energy, the drift kinetic energy 
and thermal kinetic energy of electrons, plus the growth rate and polarization of the 
EM fields, and the spectral energy distribution.

We start the simulations from the fundamental physical laws. The EM fields 
and the interaction between them and electrons can be described by Maxwell$'$s 
equations, i.e. Amp$\grave{\rm e}$re's law, Faraday's law of induction, 
Gaussian's law for magnetism, Gaussian's law and the definition of current 
\begin{equation}
\nabla\times
\textbf{B}=\mu_{0}\textbf{J}+\frac{1}{c^{2}}\frac{\partial\textbf{E}}{\partial t},
\label{eq_ampere}
\end{equation}

\begin{equation}
\nabla\times \textbf{E}=-\frac{\partial\textbf{B}}{\partial t},
\label{eq_faraday}
\end{equation}

\begin{equation}
\nabla\cdot \textbf{B}=0,
\label{eq_gaussianmag}
\end{equation}

\begin{equation}
\nabla\cdot \textbf{E}=\frac{\rho}{\epsilon_{0}},
\label{eq_gaussianele}
\end{equation}

\begin{equation}
\nabla\cdot\textbf{J}=-\frac{\partial\rho}{\partial t},
\label{eq_chargecon}
\end{equation}
where $\textbf{B}$ and $\textbf{E}$ are the magnetic field and electric field respectively, 
$\textbf{J}$ the current, $t$ the time, $c$ the speed of light, $\rho$ the charge density, 
$\epsilon_{0}$ the permittivity, $\mu_{0}$ the permeability.

The motion of electrons is governed by the Lorentz force, which can be written as 
\begin{equation}
q(\textbf{E}+\textbf{v}\times\textbf{B})=\frac{\textrm{d}m\textbf{v}}{\textrm{d}t},
\label{eq_motion}
\end{equation}
where $\textbf{v}$ is the velocity of one individual particle, $q$ the charge of a  
single particle (here it is for an electron), $m$ the electron mass. In the non-relativistic 
case, we have $m=m_{\rm e}$ where $m_{\rm e}$ is the rest mass of the electron. In the relativistic 
case, we have $m=\gamma m_{\rm e}$ where $\gamma=\frac{1}{\sqrt{1-(\textbf{v}/c)^{2}}}$ is 
the Lorentz factor. The relativistic case is also a general case for the motion equation of 
particles. 

\subsection{Initial configurations of the simulations}
\label{sec_configuration}

These equations were solved self-consistently as a pure initial value problem 
using a particle-in-cell method in a two dimensional space ($x-$ and $y-$ direction) 
and three velocity and field dimensions ($x-,~y-,~z-$ direction). Some of the numerical 
methods used here can from \citet{Omura93} and \citet{Omura05}. The Buneman-Boris method 
was used to solve the equation of 
motion \citep{Hockney81,Birdsall85}. The equation of continuity of charge was solved by 
a charge conservation method \citep{Villasenor92}. The spacing grid for the electromagnetic 
field in the present simulations is 64$\times$64. The time step in each simulation is 
$\Delta t=0.001~\nu_{\rm pe}^{-1}$ where $\nu_{\rm pe}$ is the electron plasma frequency, 
while the space step is $\Delta x= \Delta y = 0.125~\lambda _{\rm D}$ where 
$\lambda _{\rm D}=v_{\rm th}/(2\pi\nu_{\rm pe})$ is the Debye length, with $v_{\rm th}$ being 
the thermal velocity of the electrons. All the velocities in the simulations are normalized 
by the speed of light $c$. The charge-mass ratio of electron is assumed to be -1. We use an 
open boundary in the simulated system. This simulation configuration and the intrinsic 
properties of the electrons do not vary in any of the simulations. 

\begin{figure}
\centering
\includegraphics[width=9cm,clip]{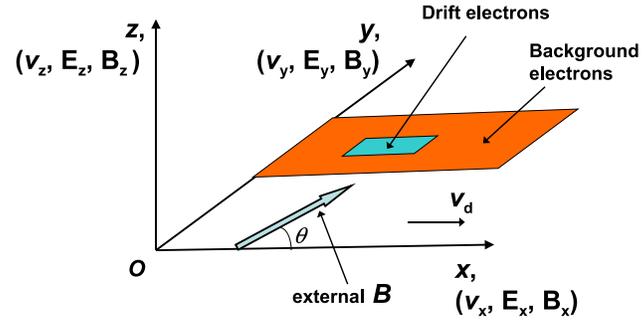}
\caption{The schematic diagram to present the simulation box for the radio emission from ultracool dwarfs. In an orthogonal 
reference system with spatial direction x, y and z, $v_{\rm x}$, $v_{\rm y}$ and $v_{\rm z}$ represent the velocity 
components of particles, while $\rm E_{x}$, $\rm E_{y}$, $\rm E_{z}$, $\rm B_{x}$, $\rm B_{y}$ and $\rm B_{z}$ represent the 
electric and magnetic fields components.  
}
\label{fig_sbox} 
\end{figure}

In each of the simulations, we assume a constant external magnetic field $\textbf{B}_{0}$ 
exists in the spatial $x-y-$plane, and the angle between $\textbf{B}_{0}$ and $x-$direction 
is defined as $\theta$ with $0^{\circ}\leqslant\theta\leqslant90^{\circ}$. The charged 
particles are initially distributed in the $x-y-$plane randomly. We assume that background 
thermal electrons may exist in the radio emission region, i.e. the magnetosphere of ultracool 
dwarfs, with number density $n_{\rm th}$ and thermal velocity $v_{\rm th}$. The injected 
electrons have a number density $n_{\rm d}$, thermal velocity $v'_{\rm th}$ and drift velocity 
$\textbf{v}_{\rm d}$ along the $x-$direction. Figure\,\ref{fig_sbox} shows the spatial simulation 
box schematically. In the simulations, we determine the strength of the external magnetic field 
via its close relation with the cyclotron frequency $\nu_{\rm ce}\approx2.8 B_{0}$ MHz. The 
relation between plasma frequency and the number of electrons is 
$\nu_{\rm pe}\approx8.98\times10^{-3}(n_{\rm e}\rm cm^{-3})^{1/2}$ MHz. 

Since we do not know the electron density in the radio emission region on ultracool dwarfs, 
a range of values and related plasma parameters are listed in Table \,\ref{tab_plasma}. 
From the values in the Table, we see that the present simulations are in a relatively 
micro-region and short timescale (0.1 ns $-$ 10 $\mu$s). 

In order to see the influence of the above parameters on the released EM waves, we set a group of 
standard values for them (see Table\,\ref{tab_parameters}). In this model, we assume the direction of the external 
magnetic field parallel to the $y-$direction. The cyclotron frequency is set to be 10 times the 
plasma frequency. We take the thermal velocity of the background electrons and the drift 
electrons as 0.01$c$. This means that the temperature of the electrons is about 
3$\times$10$^{5}$ K, determined by $T=\frac{1}{2}m_{\rm e}v_{\rm th}^{2}/k$ in the non-relativistic case, 
where $k$ is the Boltzmann constant. We take $v_{\rm d}=0.05c$. We vary the value for one of the parameters 
whilst the other parameters remain as the standard values. 
These parameters and their values are summarized in Table\,\ref{tab_parameters}. We will interpret 
these parameters in \S\,\ref{sec_iparameters}. The standard values of these parameters are derived 
from estimations of solar bursts (typically 0.1$c$ to 0.5$c$ for the drift velocity and 0.002$c$ 
to 0.05$c$ for the thermal velocity, \citet{Dulk85}), and the studies on auroral kilometric radiation on 
the Earth, Jovian millisecond bursts, and Saturnian kilometric radiation ($\sim1-10$ keV for the energetic 
electrons and $\sim$100 eV for the thermal electrons, \citet{Zarka98,Hess07a,Zarka07,Hess07b,Lamy10}). 
The values of the electron velocities in the relativistic case (see \S\,\ref{sec_relativistic}) refer to the work 
of \citet{Louarn86} and references therein. We choose the optional values over a wide range so that the approximate 
functions between energies and the parameters can be obtained. 

In this paper, we distinguish the irregular thermal motion of the electrons and their uniform motion. 
The thermal energy of the electrons in the non-relativistic case is defined and calculated from the thermal 
motion of the electrons by 
\begin{equation}
\begin{split}
E_{\rm th}&=E_{\rm tk}-E_{\rm d}\\
&=\sum_{i=1}^{n}(\frac{1}{2}m_{\rm e}\textbf{v}_{i}^{2})-\frac{1}{2}m_{\rm e}(\sum_{i=1}^{n}\textbf{v}_{i})^{2},
\end{split}
\label{eq_eth}
\end{equation}
where $n$ is $n_{\rm th}$ for the background electrons and $n_{\rm d}$ for the drift electrons, 
$m_{\rm e}$ the electron mass, $\textbf{v}_{i}$ the velocity of the $i^{th}$ particle. On the right hand side 
of this equation, the first term represents the total kinetic energy of the system $E_{\rm tk}$, and the second 
term describes the drift energy of the electrons $E_{\rm d}$. 

In the relativistic case, the energy of one electron is defined by the energy-momentum relation 
\begin{equation}
\begin{split}
E_{i \rm e}^{2}&=\textbf{p}_{i\rm e}^{2}c^{2}+m_{\rm e}^{2}c^{4}\\
&=(\gamma m_{\rm e}c^{2})^{2},
\end{split}
\label{eq_eprelation}
\end{equation}
where $\textbf{p}_{i\rm e}=\gamma m_{\rm e}\textbf{v}_{i}$ is the momentum of the $i^{th}$ electron, $\gamma=\frac{1}{\sqrt{1-(\textbf{v}_{i}/c)^{2}}}$ 
the Lorentz factor. Therefore the kinetic energy of one electron can be expressed as 
\begin{equation}
\begin{split}
E_{i \rm ek}&=E_{i \rm e}-m_{\rm e}c^{2}\\
&=(\gamma-1) m_{\rm e}c^{2}.
\end{split}
\label{eq_ek}
\end{equation}
Then the total kinetic energy of the system $E_{\rm tk}$ is 
\begin{equation}
E_{\rm tk}=\sum_{i=1}^{n} E_{i \rm ek}.
\label{eq_etk}
\end{equation}
For the drift energy of the system $E_{\rm d}$, we first define $\gamma_{\rm d}=\frac{1}{\sqrt{1-(\sum_{i=1}^{n}\textbf{v}_{i}/nc)^{2}}}$. 
Then we have 
\begin{equation}
E_{\rm d}=(\gamma_{\rm d}-1)\cdot n\cdot m_{\rm e}c^{2}.
\label{eq_ed}
\end{equation}
So the thermal energy is $E_{\rm th}=E_{\rm tk}-E_{\rm d}$. The definition of the drift energy of the system 
in both the non-relativistic and relativistic cases realises the uniform motion of the system along one direction via 
eliminating the irregular random motion of the electrons. Under these definitions, 
the non-relativistic case (i.e. Eq.\,\ref{eq_eth}) is a good approximation of the relativistic case (i.e. 
Eqs.\,\ref{eq_eprelation}, \ref{eq_ek}, \ref{eq_etk}, \ref{eq_ed}) when particles move with low 
velocities compared to the speed of light. Our calculations indicate that the equations for the relativistic case 
are valid for the velocity range $v_{i}\in[0,c)$, while the non-relativistic case is only valid for low velocity 
particles. A general definition of temperature then becomes $kT=(\gamma - 1)m_{\rm e}c^{2}$.

The energy of the EM field $w$ can be described by the Poynting theorem, so that we have 
$w=\frac{1}{2}\epsilon_{0}\textbf{E}^{2}+\frac{\textbf{B}^{2}}{2\mu_{0}}$ where we take 
$\epsilon_{0}=1$ and $\mu_{0}=\frac{1}{c^{2}\epsilon_{0}}$.
The first term on the right hand side of this equation gives the electric field energy while the second term represents 
the magnetic field energy. All kinds of energies are normalized by the initial total energy of the system to satisfy 
energy conservation. We exclude the energy of the external magnetic field $\textbf{B}_{0}$.

The growth rate is defined as 
\begin{equation}
\Gamma=\frac{\ln(E_{(t+\Delta t)}^{2})-\ln(E_{(t)}^{2})}{2\Delta t}.
\label{eq_gr}
\end{equation}
The degree of linear polarization in the EM fields $\Pi$ is 
\begin{equation}
\Pi=\frac{w_{\perp}-w_{\shortparallel}}{w_{\perp}+w_{\shortparallel}},
\label{eq_polarization}
\end{equation}
where $w_{\perp}$ is the EM field energy in the direction perpendicular to the external magnetic field, 
$w_{\shortparallel}$ the EM field energy in the direction parallel to the external magnetic field.

The background electrons are assumed to be in thermal equilibrium, so their velocity distribution obeys 
a Gaussian distribution. In this paper it is always taken as the following expression 
\begin{equation}
f_{0\rm b} = n_{\rm th}{\rm exp}(-\frac{v_{x}^{2}+v_{y}^{2}+v_{z}^{2}}{v_{\rm th}^{2}}),
\label{eq_vdisb}
\end{equation}
where $v_{x},~v_{y},~v_{z}$ are the components of the velocity $\textbf{v}_{\rm b}$ 
of the background electrons along $x-,y-,z-$ direction respectively.

The injected electrons have an intrinsic thermal velocity distribution but due to the acceleration they 
will obtain a drift velocity along some direction. For the purpose of simplicity, we assume all of the 
injected electrons are accelerated along one direction. We take the direction as the $x$-direction in 
our simulations, so that the initial velocity distribution for the injected electrons can be expressed as  
\begin{equation}
f_{0\rm d} = n_{\rm d}(\frac{v'_{x}-v_{\rm d}}{v'_{\rm th}})^{2l}{\rm exp}(-\frac{(v'_{x}-v_{\rm d})^{2}+v_{y}^{'2}+v_{z}^{'2}}{v_{\rm th}^{'2}}),
\label{eq_vdisd}
\end{equation}
where $l=0,1,2,3,...$, and $v_{x}^{'},~v_{y}^{'},~v_{z}^{'}$ are the components of the velocity $\textbf{v}_{\rm in}$ 
of the injected electrons along the $x-,y-,z-$ direction respectively. $l$ is the parameter which describes the size of the 
loss-cone in velocity space. In the present simulations, we take $l=0$ and $3$. All the particles are assumed 
to be randomly distributed in space.

\section{Results I: $l$=0}
\label{sec_resultsI}

When $l=0$, Eq.\,\ref{eq_vdisd} becomes a Gaussian distribution. In this section, we investigate the 
influence of the injected electrons with the Gaussian velocity distribution on the evolution of the space and 
velocity distribution of the electrons, the energy conversion efficiency, the transfer between drift kinetic 
energy and thermal energy of the system, and the growth rate and polarization of the released EM waves.

\subsection{Standard model for beam-driven ECM}
\label{sec_smodel}

We first present the simulation with standard parameters in detail; 
this we call the standard model. This standard 
configuration means that the background electrons and the injected electrons 
have the same temperature (10$^{5}$ K) but the density of the latter is higher. 
(Note we use the definition of the temperature in \S\,\ref{sec_configuration} which 
reflects the measure of the velocity dispersion). 
This is for the purpose of 
modelling a dense electron beam injected into the 
magnetosphere of an ultracool dwarf. 

\begin{figure*}
\centering
\includegraphics[width=17cm,clip]{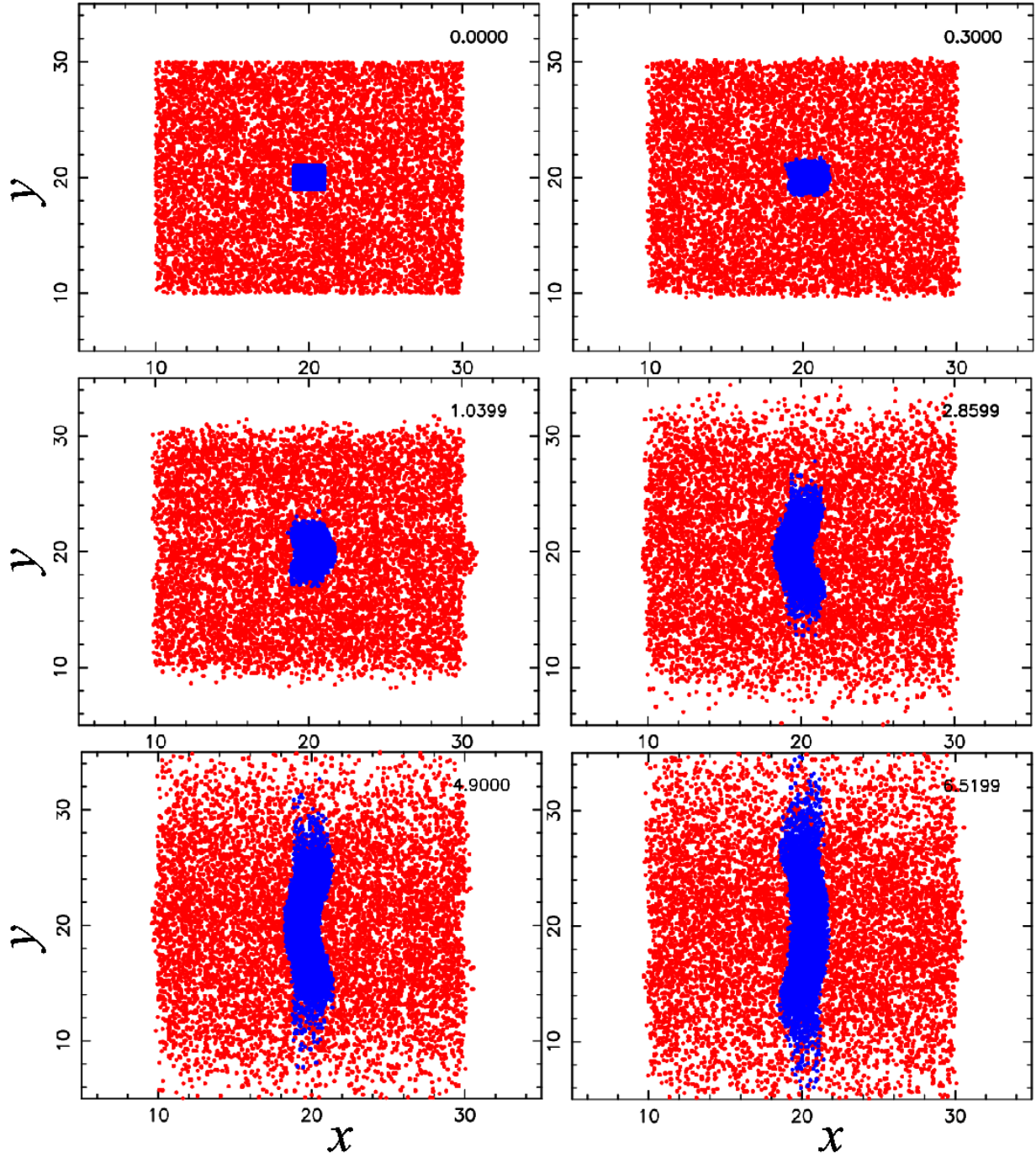}
\caption{The history of the distribution of the background electrons (red points) and beam electrons (blue points) in spatial x-y-plane. 
Different panels from top-left to bottom-right represent the snapshots for time $t=0.00$, 0.30, 1.04, 2.86, 4.90 and 6.52 $\nu_{\rm pe}^{-1}$. }
\label{fig_space1} 
\end{figure*}

Figure\,\ref{fig_space1} shows the spatial evolution of the electrons (A movie is available 
for the spatial evolution of the system at URL: http://www.arm.ac.uk/$\sim$syu/2decm/beam/beam\_space/). 
The time in each snapshot is $t=0$, $t=0.3$, $t=1.04$, $t=2.86$, $t=4.9$, 
$t=6.52~\nu_{\rm pe}^{-1}$ from the top-left to bottom-right. Due to 
the existence of the external magnetic field, the motion of the electrons 
along the $x-$direction is confined and they can only freely move along 
the $y-$direction which is parallel to the magnetic field. The motion 
of the injected electrons as a whole should be in a helical orbit with 
the radius determined by $r_{\rm d}=\frac{m_{\rm e}v_{\rm d}}{qB}$ since 
the induced magnetic field is very small. 
In this simulation, $r_{\rm d}$ is about 0.715. Since we only perform 
a 2D simulation, we see the oscillation of the injected electrons in 
the $x-y$-plane instead of a helical motion. 

The mix of background electrons and injected electrons in the 
velocity space triggered by the EM field is 
rather interesting. Fig.\,\ref{fig_velocity1} shows snapshots of the 
velocity distribution of the electrons at the same time as in Fig.\,\ref{fig_space1} 
(A movie is available for the velocity evolution at URL: http://www.arm.ac.uk/$\sim$syu/2decm/beam/beam\_velocity/).  
As seen in the top-left panel in the figure, the velocities of 
both background electrons and injected electrons are initially  
a Gaussian distribution, with the injected electrons 
having a drift velocity of $0.05c$. When time evolves, the current 
generated by the injected electrons induces a strong 
electric field along the direction perpendicular to the external 
$\textbf{B}_{0}$ which only alters the direction of the 
flow. This electric field accelerates a fraction of the background 
electrons so that a tail at high velocity is 
developed in $v_{\rm x}$ and $v_{\rm z}$ (see the velocity distribution 
in each time snapshot), whilst the injected 
electrons are decelerated by the electric field, losing their 
drift velocity along the $x-$direction gradually and wavily. Also 
because of the perpendicular electric field, the injected electrons gain 
a drift velocity in $v_{\rm z}$ which oscillates 
between $-0.1c$ and $0.1c$. The appearance of a double peak in the 
distribution of $v_{\rm y}$ is due to the magnetic constraint and the acceleration 
of the induced electric field on the 
perpendicular velocity $v_{\rm x}-v_{\rm y}$. One can imagine that 
if the evolutionary time is sufficiently long the electrons may separate into 
two groups. One group having a velocity along the direction of the external 
$\textbf{B}_{0}$, while the other group has the opposite velocity direction.

The consequence of this process is that the velocities of the electrons 
evolve from a concentrated Gaussian distribution  
to an expanded quasi-Gaussian distribution with a high velocity tail. 
During the diffusion process of the particles in 
velocity space, an energy transfer occurs between the kinetic energy of 
the electrons and the induced EM field energy. As 
the injected electrons move in the external $\textbf{B}_{0}$ at time $t=0$, they 
start releasing their kinetic energy to the EM wave 
energy in the manner of an increase in the induced EM field strength. 
After some time, the EM field energy will reach its maximum whilst  
$E_{\rm tk}$ approaches a minimum. Then the EM 
field energy may be absorbed by the electrons to compensate for lost 
kinetic energy. The $E_{\rm tk}$ will increase after 
a short time as the field energy decreases. 
Oscillations in $E_{\rm tk}$ and the field energy will  
last for some time until the system is balanced and there is an anti-phase 
relation between the two kind of energies. 

Figure\,\ref{fig_senergy} illustrates the evolution of the total kinetic 
energy $E_{\rm tk}$, drift energy $E_{\rm d}$ 
and thermal energy $E_{\rm th}$ of the electrons and the field energy $w$. 
We plot the same time points for the space and 
velocity distribution of the electrons in this figure as solid black circles. 

As we can see from this figure, $E_{\rm tk}$ and $w$ are exactly anti-phase 
as expected. The multi-peaks of $w$ should 
be associated with the different wave mode in frequency space which will be 
shown later. The relation of the fine structure 
of $E_{\rm d}$ and $E_{\rm th}$ is not as obvious as the relation between 
$E_{\rm tk}$ and $w$ since the interaction 
between $E_{\rm d}$ and $E_{\rm th}$ is via the EM field as a time and 
space delay can affect the phase relation. However, 
we see that $E_{\rm d}$ decreases dramatically at the starting time when 
$E_{\rm th}$ increases rapidly. After about 
1.5 $\nu_{\rm pe}^{-1}$, when $E_{\rm d}$ and $E_{\rm th}$ approximately 
have equal values, the decrease of $E_{\rm d}$ 
slows, and so does the increase in $E_{\rm th}$. At time 7.1 
$\nu_{\rm pe}^{-1}$, at least 70\% of $E_{\rm d}$ is converted to 
$E_{\rm th}$. This is interesting because it means that the transverse motion 
of electrons in an external magnetic field may 
be an efficient way to heat the ambient electrons. 

The maximum growth rate of the EM wave in this simulation is about 
9.68$\times10^{2}~\nu_{\rm pe}$. The polarization of the released EM waves is 
highly linear or circular, depending on the initial configuration. 
More details on the growth rate and polarization in the standard model will 
be shown in the next section. Figure\,\ref{fig_spolarization} shows the evolution 
history of the EM waves energy perpendicular and parallel to the external magnetic 
field. Again, we use the solid black circles to denote the time points for the space 
and velocity distribution of the electrons shown in Fig.\,\ref{fig_space1} and \ref{fig_velocity1}. 
From this figure, we clearly see that most of the EM waves are polarized in the 
direction perpendicular to $\textbf{B}_{0}$, whilst only a very 
small fraction of the EM waves is released parallel to $\textbf{B}_{0}$ 
($\sim$10$^{-3}$ of the perpendicular energy).

\subsection{Influences of parameters}
\label{sec_iparameters}

In this section, we investigate the influence of the following parameters on the energy history, 
growth rate and polarization of the EM waves.
\\
{\it (a) The strength of the external magnetic field $B_{0}$} 
\\
$B_{0}$ can be determined by the cyclotron frequency
\begin{equation}
B_{0}\approx 0.357\times10^{3}\frac{\nu_{\rm ce}}{{\rm GHz}}~~~~\rm Gauss.
\end{equation}
To date, all detected radio emission of ultracool dwarfs are in the GHz band, while observations 
performed with the NRAO very large array in 2007 show no trace of radio emission from two UCDs at 325 MHz, 
placing an upper flux limit of $\sim 900$ $\mu$Jy at 2.5$\sigma$ level \citep{Jaeger11}. From this, we infer 
that the magnitude of $B_{0}$ is few kilo-Gauss (if the radio emission is truly at the cyclotron frequency). 
Here, we take $\nu_{\rm ce}/\nu_{\rm pe}=0,~5$ and 10 to see how the magnetic field can affect the simulation results. 
When $\nu_{\rm ce}/\nu_{\rm pe}=0$, there is no external magnetic field.
\\
{\it (b) The angle $\theta$ between the magnetic field and the drift velocity}
\\
$\theta$ is one of the crucial parameters to influence the direction of the radiated EM waves, 
i.e. the polarization. When $\theta=0^{\circ}$, the injected electrons move uniformly parallel 
to the external $\textbf{B}_{0}$ in addition to the irregular thermal motion. When $\theta=90^{\circ}$, 
the motion of the injected electrons is perpendicular to $\textbf{B}_{0}$.
\\
{\it (c) The drift velocity $v_{\rm d}$}
\\
We expect that the radio emission may be enhanced by increasing the value of $v_{\rm d}$. 
In this section, we take $v_{\rm d}=0c$, $0.005c$ and $0.01c$ to avoid the relativistic 
effect where gyrosychrotron emission plays an important role. When $v_{\rm d}=0c$, we see 
the effect of pure thermal electrons moving in the external $\textbf{B}_{0}$ on the induced EM waves.
\\
{\it (d) The temperature $T$}
\\
We investigate the response of the radiated EM waves and the transfer between the energies by varying 
$v_{\rm th}$. We take $v_{\rm th}=0.005c$, $0.01c$, and $0.05c$ for which the corresponding 
$T$ is $7.5\times10^{4}$, $3\times10^{5}$, $7.5\times10^{6}$ K.
\\
{\it (e) The background electrons by changing $n_{\rm th}/n_{\rm d}$}
\\
Since our computation capability is limited, we only investigate the effect of the existence of the 
background electrons on the EM waves and the energy transport. We take $n_{\rm th}/n_{\rm d}=0,1,2$.  
When $n_{\rm th}/n_{\rm d}=0$, there are no background electrons. 

\begin{figure*}
\centering
\includegraphics[width=11cm,clip,angle=90]{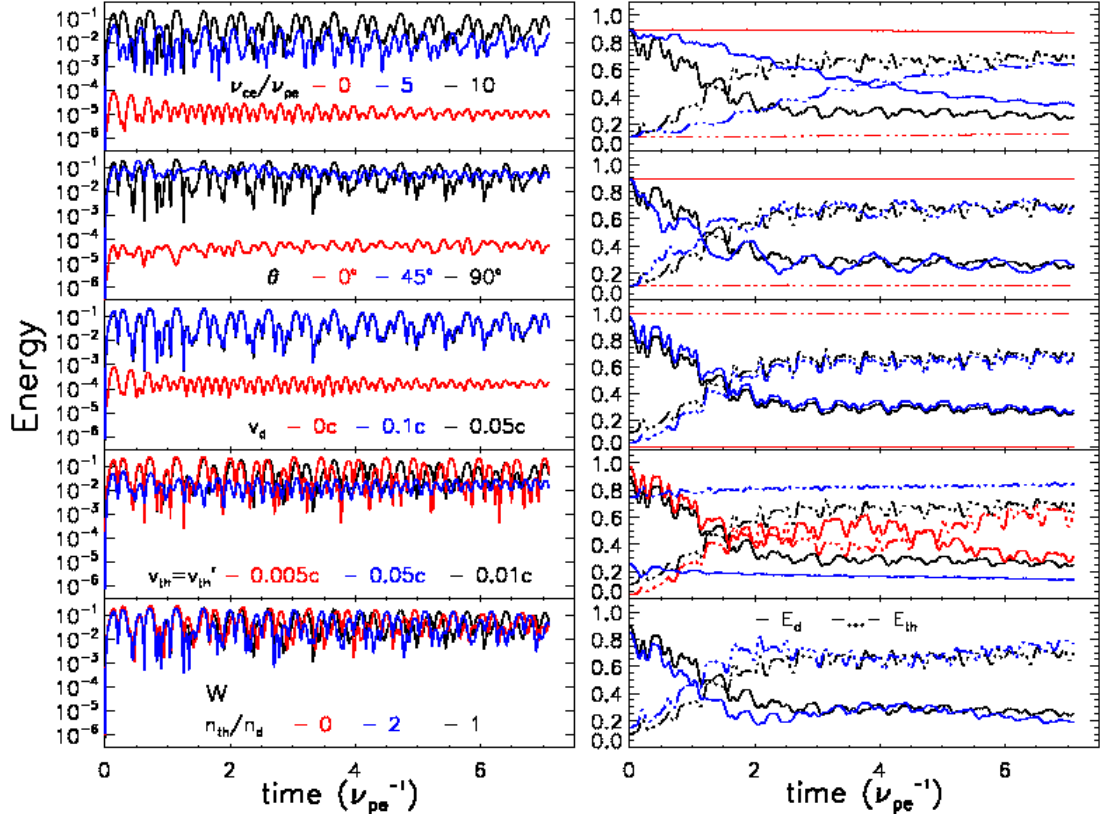}
\caption{The evolution of the field energy (left panels), total thermal energy and the drift energy (right panels). 
Different panels are for different parameters, i.e. 1$^{st}$ row - $\nu_{\rm ce}/\nu_{\rm pe}=0$ (red line),
5 (blue line), 10 (black line); 2$^{nd}$ row - $\theta=$0$^{\circ}$ (red line), 45$^{\circ}$ (blue line),90$^{\circ}$ (black line); 
3$^{rd}$ row - $v_{\rm d}=$0$c$ (red line), 0.1$c$ (blue line), 0.05$c$ (black line); 4$^{th}$ row - $v_{\rm th}=$0.005$c$ (red line),
0.05$c$ (blue line), 0.01$c$ (black line); 5$^{th}$ row - $n_{\rm th}/n_{\rm d}=$0 (red line), 2 (green line), 1 (black line). 
In the right panels, the solid line is the drift energy while the dot-dot-dot-dashed lines is for the thermal energy. Note that 
the black colored lines in this figure are always for the model with standard parameters. }
\label{fig_genergy} 
\end{figure*}

Figure\,\ref{fig_genergy} shows the energy evolution in each simulation. We see that the 
EM field energy (left panels) are in a very low level in all three specific cases (see the colored lines). In 
the first case, there is no magnetic field (which can be understood as the injected electrons just 
pass by very quickly without sufficient energy transfer via the EM field). In the second case, the 
motion of the injected electrons is parallel to the external magnetic field. In this case, the induced 
electrons can not remain within the simulation box since the magnetic field can not constrain them. In 
the third case, the drift velocity of the injected electrons is 0$c$; although the injected electrons 
can remain and interact with the background, there is no coherent current, i.e. the field energy is 
still small although higher than in the other two cases.

A common expression of these cases is $\textbf{v}\times\textbf{B}_{0}=0$ which is the Lorentz force induced by 
the external magnetic field. This force makes the initially coherent current bend (i.e. the electrons with 
drift velocity), leading to spatial curled EM fields that is the medium to accomplish energy transport 
among different kind of energies. 

In other cases except the above three cases, the field energy can maintain a much higher level after they 
reach the maximum, typically orders of $2-3$ that of the above cases. As shown in Fig.\,\ref{fig_genergy}, 
the field energies in all cases oscillate with large amplitude caused by the transfer between the kinetic energy 
and field energy. In other words, these oscillations reflect the emission and absorption of electrons to the EM 
waves. With the achievement of the diffusion process of the electrons in the velocity space, a dynamic balance is  
approached. This results in a gradual decrease in the amplitudes of the oscillations with time. The time 
for relaxation of the field energy is $>$10 $\nu_{\rm pe}^{-1}$ which is much longer than the time 
taken for the field energy to reach maximum, $\sim$0.14 $\nu_{\rm pe}^{-1}$. 

\begin{figure*}
\centering
\includegraphics[width=11cm,clip,angle=90]{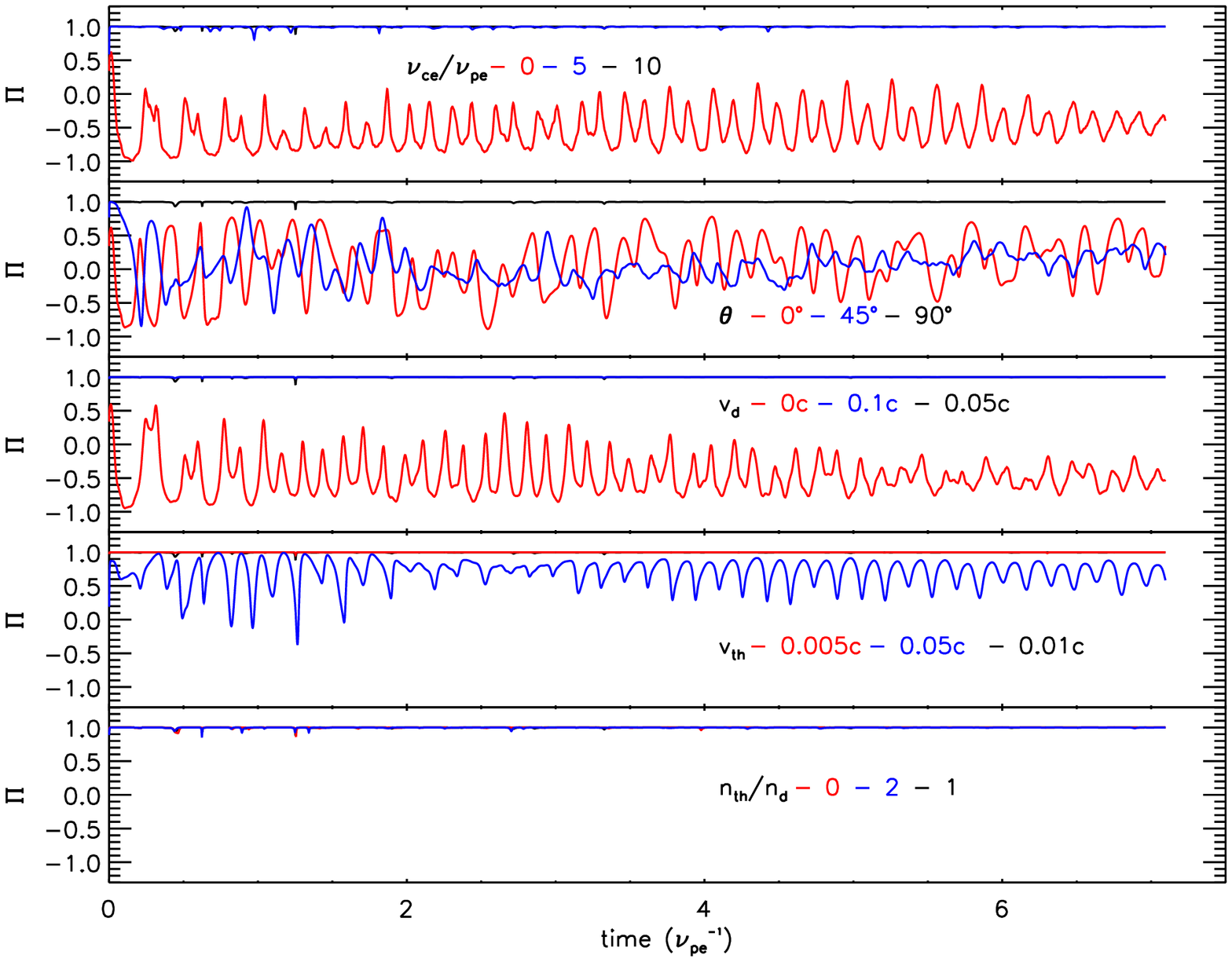}
\caption{The time evolution of the degree of linear polarization in the simulations. Each panel is for one parameter, 
i.e. from top to bottom, $\nu_{\rm ce}/\nu_{\rm pe}$, $\theta$, $v_{\rm d}$, $v_{\rm th}$, $n_{\rm th}/n_{\rm d}$. Different 
colors in each panel are for different values of the parameters as shown in the panel. }
\label{fig_gpolarization} 
\end{figure*}

As expected, increasing $B_{0}$ by a factor of two, i.e. $\nu_{\rm ce}/\nu_{\rm pe}$ from 5 to 10, leads a rise 
in the field energy. It seems that the mean value of the field energy is not sensitive to 
$\theta$ when $\theta=45^{\circ}$ and $90^{\circ}$. However, the modes (or direction) of the EM waves are affected. 
In the case of $\theta=45^{\circ}$, the EM waves have a similar energy level in the direction  
perpendicular and parallel to $\textbf{B}_{0}$, which is shown in Fig.\,\ref{fig_gpolarization} where  
we discuss polarization. The increase of $v_{\rm d}$ from $0.05c$ to $0.1c$ only rises the energy level slightly 
in the present simulations. In fact, when $v_{\rm d}$ is sufficiently high, we have to consider the relativistic effect 
and hence gyrosynchrotron radiation which will be addressed in \S\,\ref{sec_relativistic}. The mean energy level of the EM 
waves is not sensitive to the thermal velocity and background electron number density. Note that in the standard model, 
the initial density of the background electrons is $\sim 100$ times less than that of the injected electrons. 

The polarization of the EM waves, or their energy distribution with respect to the magnetic field direction, is shown in 
Fig.\,\ref{fig_gpolarization} quantitatively. The EM waves in the standard model are 100\% linearly polarized. 
When $\textbf{v}\times\textbf{B}_{0}=0$, the EM waves frequently switch their direction from parallel-dominant to 
perpendicular-dominant, and rarely do they have up to 50\% linear polarization. This behaviour indicates that 
the waves have a significant linearly-polarized component. In addition to $\theta$, thermal 
motion (i.e. temperature) of the electrons can influence the level of polarization. As we see from the bottom 
second panel in Fig.\,\ref{fig_gpolarization}, the electrons with temperature $7.5\times10^{6}$ K can generate 
the EM waves with $\sim$50\% $-$ $\sim$90\% linear polarization, while 100\% linearly polarized waves are  
generated by the electrons with temperature $7.5\times10^{4}$ and $3\times10^{5}$ K. 
A low density of background electrons does not alter the highly linear polarization in the present simulations.

The dissipation of the drift energy in the injected electrons (when they move in the external magnetic field) is important 
because this may be sufficient to increase the thermal energy of the system. We show the history of the drift 
energy and thermal energy in Fig.\,\ref{fig_genergy}. Comparing the right panels in Fig.\,\ref{fig_genergy}, we 
find that: (i) when $\textbf{v}\times\textbf{B}_{0}=0$, the energy transfer is the least efficient. In this case, 
there is almost no energy exchange; (ii) when initially $\textbf{v}\times\textbf{B}_{0}\neq 0$ and $E_{\rm d}>E_{\rm th}$, 
$E_{\rm d}$ can be transported to $E_{\rm th}$ rapidly; the timescale in which $E_{\rm d}$ and $E_{\rm th}$ reach the same value 
is about $\sim1 - 5$ $\nu_{\rm pe}^{-1}$.

\begin{figure}
\centering
\includegraphics[width=7cm,clip,angle=90]{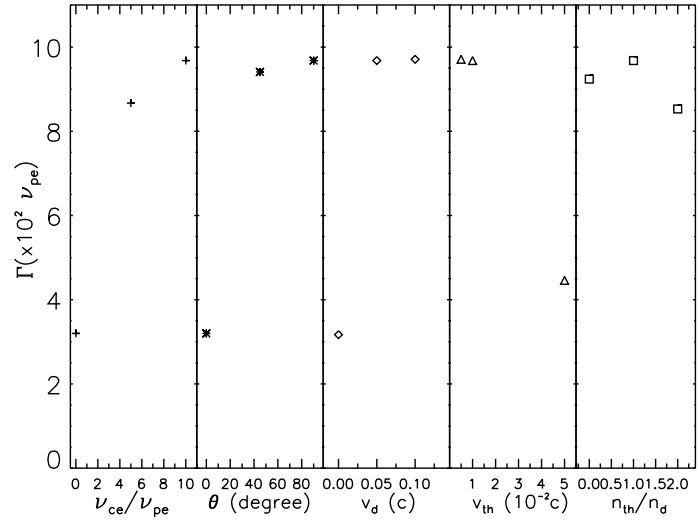}
\caption{The maximum growth rate of the field energy in each simulation. }
\label{fig_ggr} 
\end{figure}

Figure\,\ref{fig_ggr} illustrates the growth rate of the EM waves in the simulations. 
The increase of the system temperature ($3\times10^{5}-7.5\times10^{6}$ K) 
will suppress the growth rate significantly. The number of the electrons also affects the growth rate, but their relation is 
not so obvious. 

\subsection{Spectrum}
\label{sec_spectrum}

\begin{figure*}
\centering
\includegraphics[width=11cm,clip,angle=90]{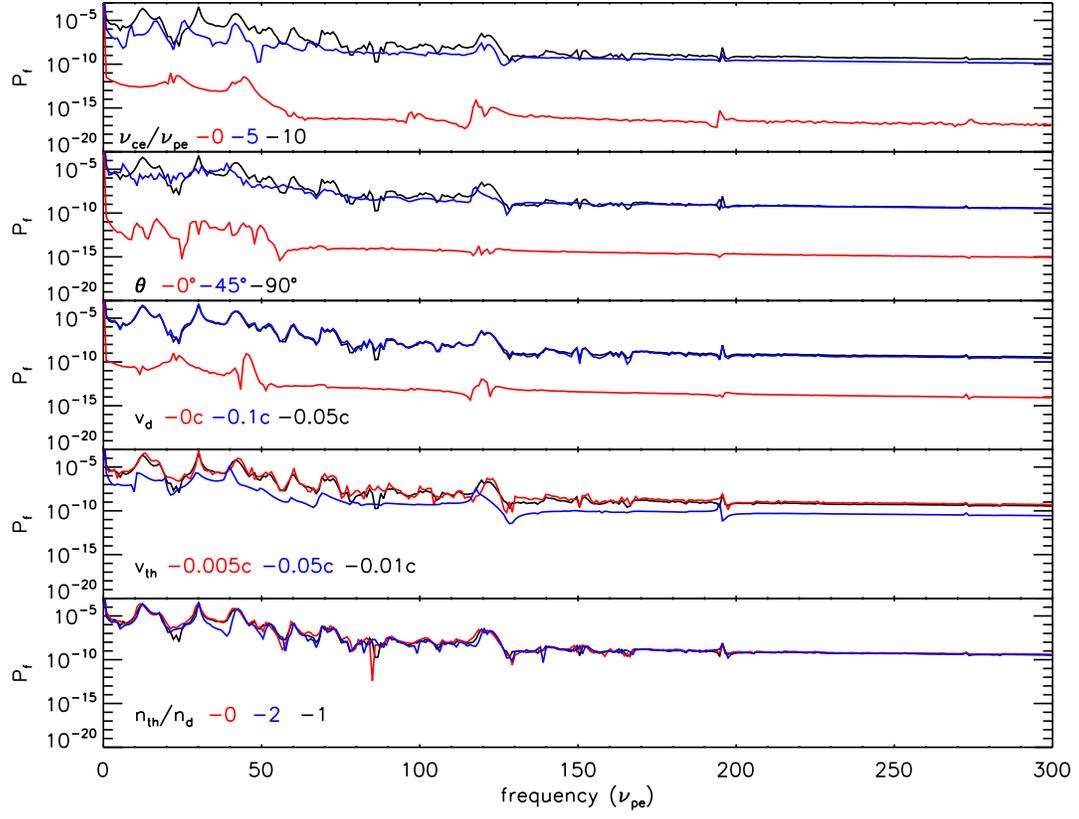}
\caption{Same as Fig.\,\ref{fig_gpolarization} but for the spectral energy density ($\rm P_{f}$) in each of the simulations 
in the non-relativistic beam driven instability.}
\label{fig_gspectrum} 
\end{figure*}

The EM field (wave) energy is the integral (or sum if the signal is discrete) of the contribution from different frequencies. 
In order to obtain the energy distribution of the EM field in frequency space (i.e. the spectral energy 
distribution, SED), we perform a fast Fourier transform (FFT) to the EM field energy history. 
Figure\,\ref{fig_gspectrum} illustrates the SED. 

We clearly see many emission and absorption lines in Fig.\,\ref{fig_gspectrum} which represent the EM field energy distribution 
at different frequencies. We find the majority of the field energy is from frequencies $<80\nu_{\rm pe}$ with bandwidth 
at half maxima $\sim6\nu_{\rm pe}$. 

\section{Results II: $l=3$}
\label{sec_resultsII}
In order to see the interactions of the electrons with different initial velocity distributions, we perform 
another series of simulations with $l=3$. 
In this case, when $v_{\rm d}=0$, Eq.\,\ref{eq_vdisd} is a typical loss-cone velocity distribution. 
Note that in this section, we take $v_{\rm d}=0$ in the standard model to exclude the effect of the electron-beam. 
We vary both the thermal velocity of the electrons to see the effect of the temperature on the energy exchange and 
release (Movies are available for the spatial evolution of the system 
at URL: http://www.arm.ac.uk/$\sim$syu/2decm/losscone/\\
losscone\_l=3\_space/ and for the   
velocity evolution URL: http://www.arm.ac.uk/$\sim$syu/2decm/losscone/\\
losscone\_l=3\_velocity/). 

\subsection{Standard model and influence of parameters in loss-cone-driven ECM}
\label{sec_lsmodel}

\begin{figure*}
\centering
\includegraphics[width=11cm,clip,angle=90]{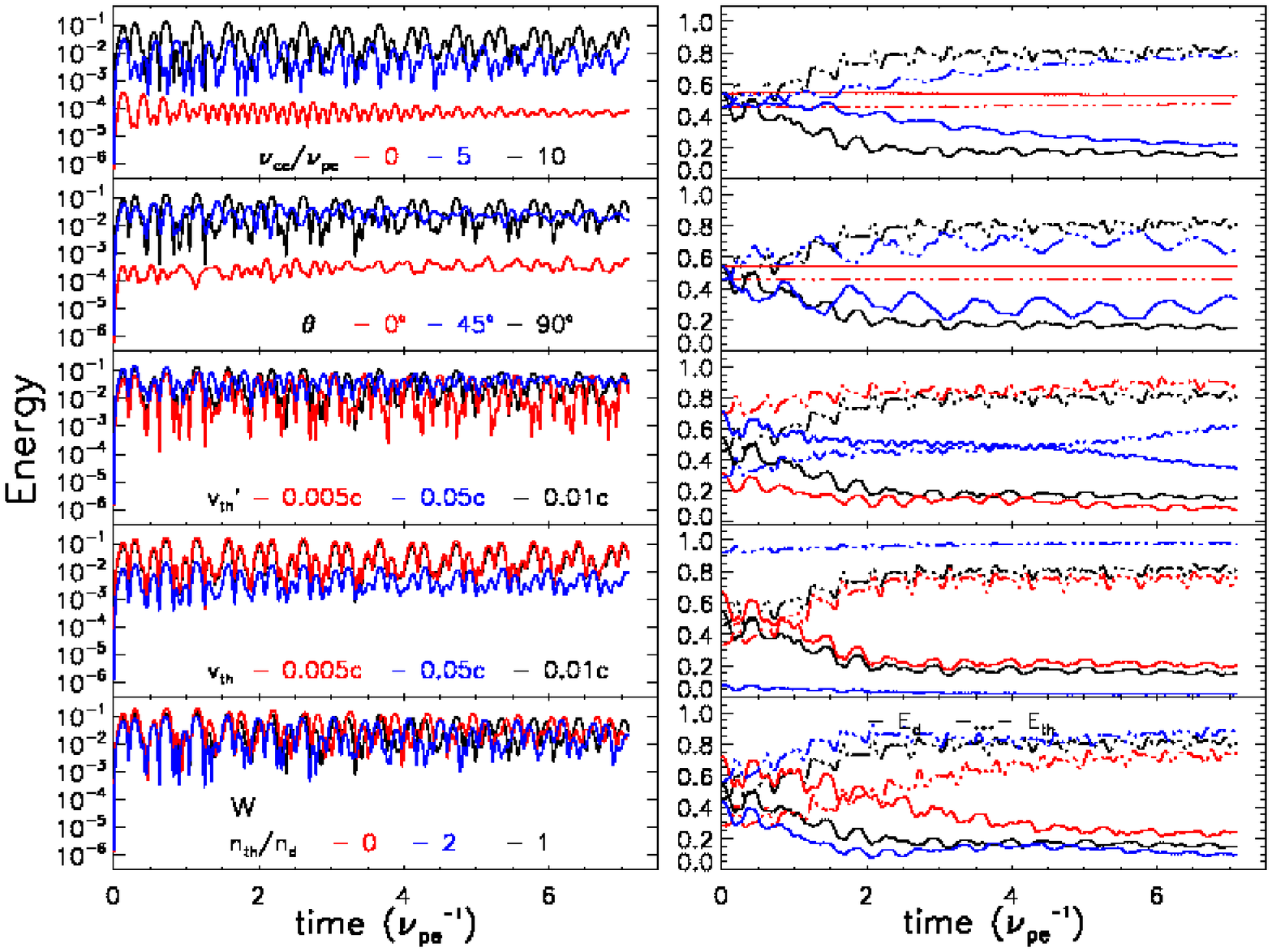}
\caption{Same as Fig.\,\ref{fig_genergy} but for the loss-cone-driven ECM. }
\label{fig_lenergy} 
\end{figure*}

Figure\,\ref{fig_lenergy} illustrates the EM field in the left panels and in the right panel we have the thermal and drift 
energy of the electrons from the loss-cone driven ECM. From this figure, we see that 
only a very small fraction of kinetic energy is converted to EM field 
energy if there is no external magnetic field or the loss-cone velocity distribution is along the direction parallel to the 
external magnetic field.  

\begin{figure}
\centering
\includegraphics[width=7cm,clip,angle=90]{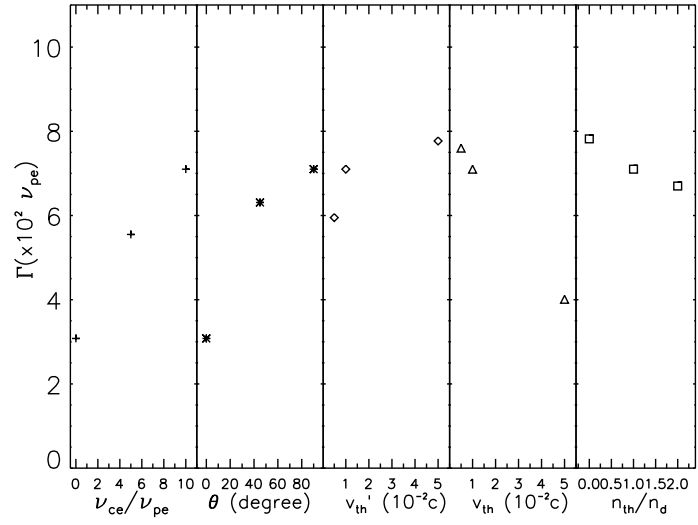}
\caption{The maximum growth rate of the field energy in the loss-cone-driven ECM. }
\label{fig_lgr} 
\end{figure}

In other cases, an increase in any one of the parameters $B_{0}$, $\theta$ and $v'_{\rm th}$ leads to an increase in the 
induced EM field energy. Rising the temperature of the background electrons (i.e. the thermal level) can suppress the 
growth of the field energy. Since we only vary the number of background electrons in a very small range, it does not 
affect the field energy significantly. As seen from the left-bottom panel in Fig.\,\ref{fig_lenergy}, increasing the number of background 
electrons decreases the induced EM field energy. These results are consistent 
with the growth rate of the EM wave illustrated in Fig.\,\ref{fig_lgr}. From the right panels in Fig.\,\ref{fig_lenergy}, 
we see that the drift energies are rapidly dissipated, eventually leading to irregular motion of the electrons in the external 
magnetic field. 

\begin{figure*}
\centering
\includegraphics[width=11cm,clip,angle=90]{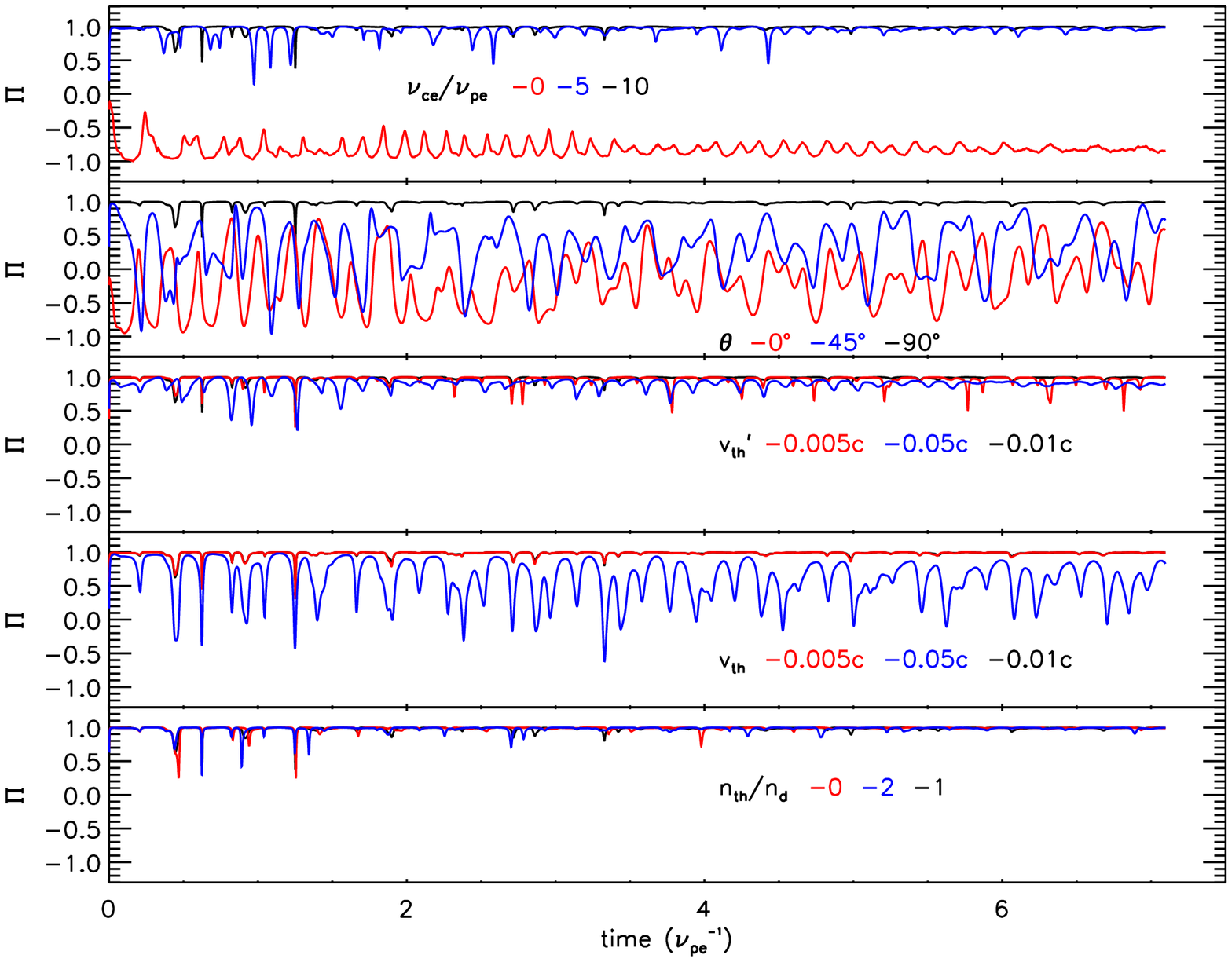}
\caption{The time evolution of the degree of linear polarization in the loss-cone-driven ECM. 
Each panel is for one parameter, i.e. from top to bottom, 
$\nu_{\rm ce}/\nu_{\rm pe}$, $\theta$, $v_{\rm d}$, $v_{\rm th}$, $n_{\rm th}/n_{\rm d}$. Different 
colors in each panel are for different values of the parameters as shown in the panel. }
\label{fig_lpolarization} 
\end{figure*}

The history of the degree of the polarization of the EM waves in each simulation is shown in Fig.\,\ref{fig_lpolarization}. 
The conditions for the circularly polarized EM waves are notable, ranging from (1) decreasing the magnetic field; (2) varying the angle 
$\theta$ from perpendicular to non-perpendicular; to (3) increasing the thermal level of the background electrons. 
We suggest these conditions may be associated with the circularly polarized components of the radio emission from 
ultracool dwarfs. 

\subsection{Spectrum}
\label{sec_lspectrum}

\begin{figure*}
\centering
\includegraphics[width=11cm,clip,angle=90]{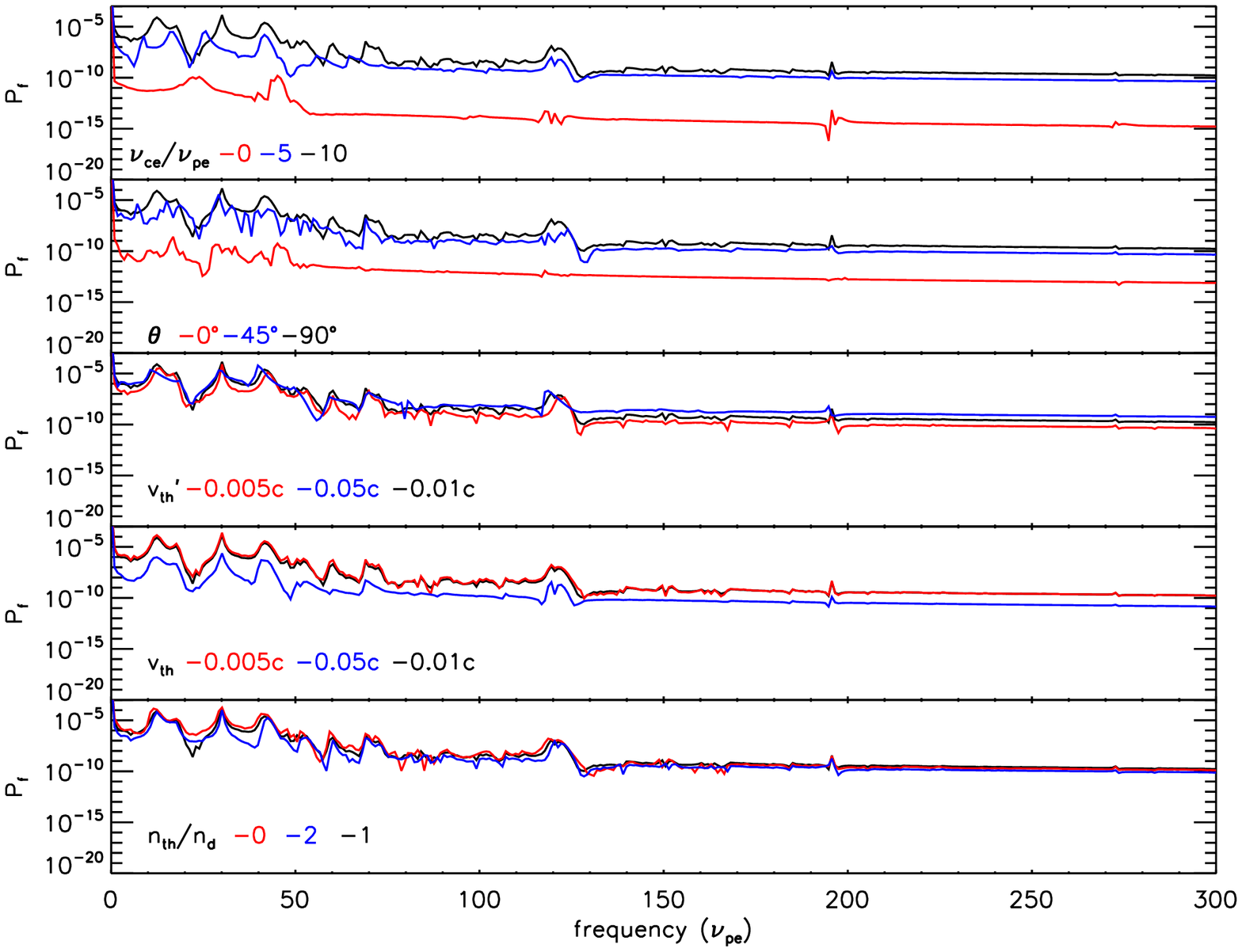}
\caption{Same as Fig.\,\ref{fig_gspectrum} but for the loss-cone-driven ECM.}
\label{fig_lspectrum} 
\end{figure*}

The influence of the parameters on the spectrum is shown in Fig.\,\ref{fig_lspectrum} in which we find that 
the magnetic field and the angel $\theta$ can affect the spectral energy distribution significantly, while the 
thermal effect and the number of background electrons only plays a minor role. Some frequency bands are notable, 
e.g. from 10 to 70$\nu_{\rm pe}$.

\subsection{Comparison between beam-driven and loss-cone-driven ECM}

Comparison between the different initial velocity distributions can help us understand the roles of the initial 
parameters in the process of releasing EM wave energy and the transfer between drift energy and thermal energy. Combining 
Fig.\,\ref{fig_genergy} \& \ref{fig_gpolarization} (beam-driven) and Fig.\,\ref{fig_lenergy} \& \ref{fig_lpolarization} 
(loss-cone-driven), we see that:\\
$~~$ 1) the existence of the drift kinetic energy of the electrons can be considered 
as a coherent current (which is necessary to generate the intense EM field energy), while the form of the initial velocity distribution 
is not important;\\
$~~$ 2) in order to efficiently obtain the intense EM field energy, the external magnetic field plays a crucial role. The 
angle between the magnetic field and the coherent current significantly affects the strength of the released EM field energy 
in a non-linear relation (see Fig.\,\ref{fig_ggr} and \ref{fig_lgr}) and also the propagation direction of the EM waves;\\
$~~$ 3) pure thermal motion of the injected electrons in the external magnetic field may play a role in generating the EM waves (e.g. 
when $v_{\rm d}=0$ in the beam-driven ECM, middle panel in Fig.\,\ref{fig_genergy}), while the thermal level of the background 
electrons mainly suppress the generation of the EM waves. 

The SED of the beam-driven and loss-cone-driven ECM indicate that all the parameters (except 
the number of background electrons in the present simulations) can affect the SED, however certain harmonic 
frequency bands will appear if the coherent current is sufficiently strong, for example see 
the region from 10 to 70$\nu_{\rm pe}$. There is a negligible signal in the very high frequency band 
$>100\nu_{\rm pe}$. Also, it seems that the SED weakly depends on the number of background electrons, 
but this needs further investigation since we do not vary the number of particles over a sufficiently 
wide range in the simulations.

\section{Relativistic beam-driven instability}
\label{sec_relativistic}

In this section, we show the case where a relativistic electron-beam moves in an external magnetic field. 
We take $l=0$ in Eq.\,\ref{eq_vdisd} and set the same standard parameter values as in \S\,\ref{sec_resultsI} except here we take  
a larger value for the drift velocity of the electrons, i.e. $v_{\rm d}=0.98c$ for the standard model; $0.58c$ and 
$0.78c$ for the optional values. Furthermore, we only vary the thermal velocity of the background electrons, with 
values of $v_{\rm th}=0.001c$, $0.01c$ (standard value), and $0.1c$.  

\subsection{Influences of parameters}
\label{sec_rmodel}

\begin{figure*}
\centering
\includegraphics[width=11cm,clip,angle=90]{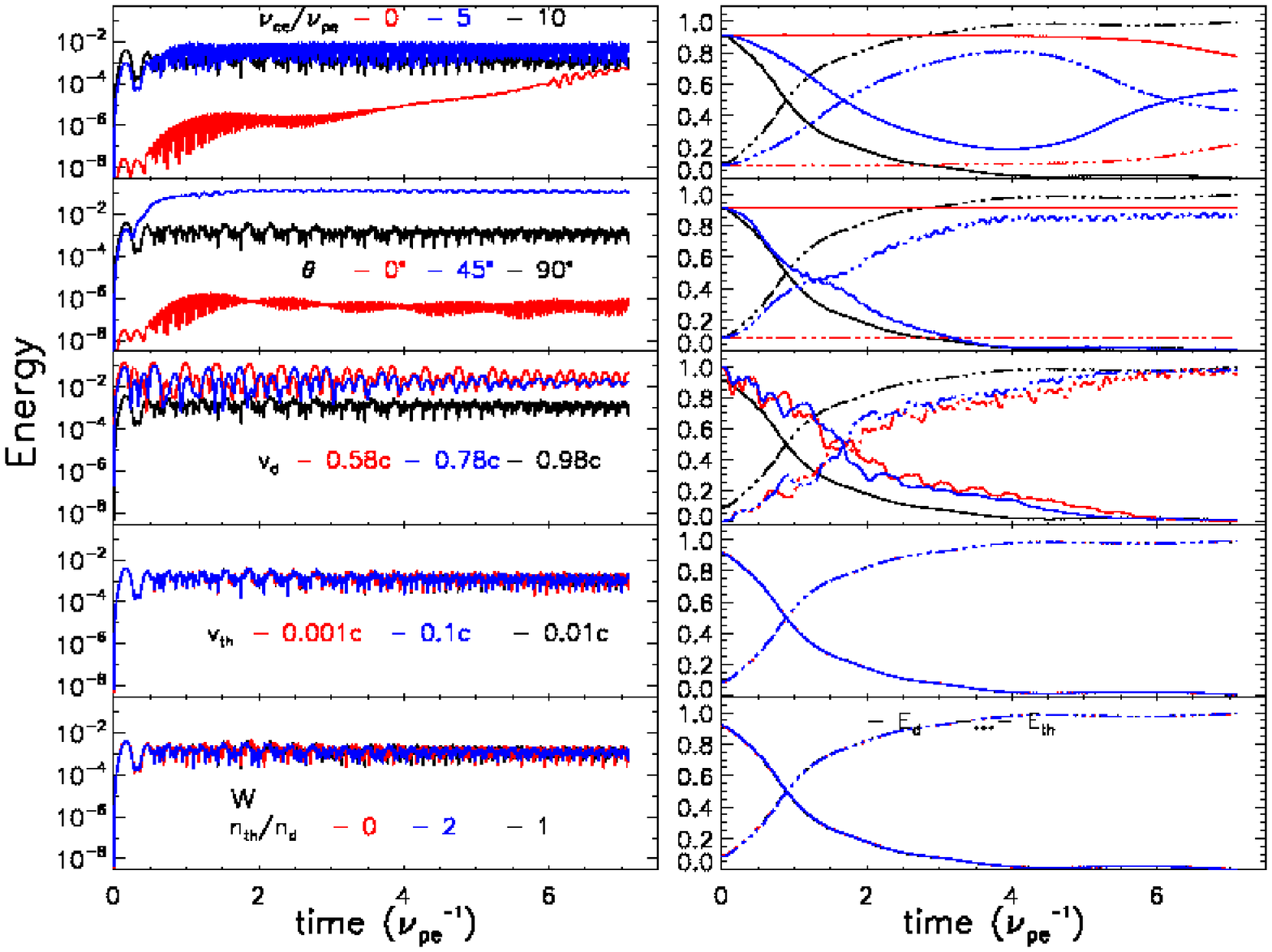}
\caption{Same as Fig.\,\ref{fig_genergy} but for the relativistic beam-driven instability. }
\label{fig_renergy} 
\end{figure*}

\begin{figure*}
\centering
\includegraphics[width=11cm,clip,angle=90]{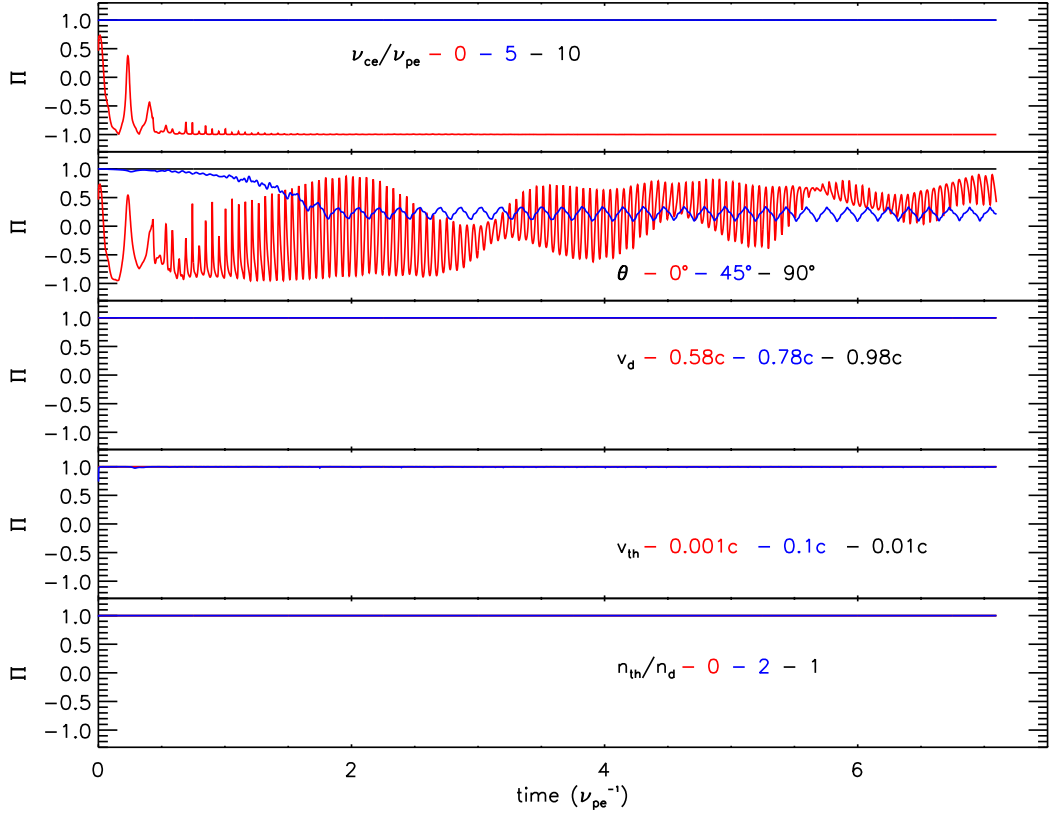}
\caption{The evolution of the degree of linear polarization with time in the relativistic beam-driven instability. 
Each panel is for one parameter, i.e. from top to bottom, 
$\nu_{\rm ce}/\nu_{\rm pe}$, $\theta$, $v_{\rm d}$, $v_{\rm th}$, $n_{\rm th}/n_{\rm d}$. Different 
colors in each panel are for different values of the parameters as shown in the panel. }
\label{fig_rpolarization} 
\end{figure*}

\begin{figure}
\centering
\includegraphics[width=7cm,clip,angle=90]{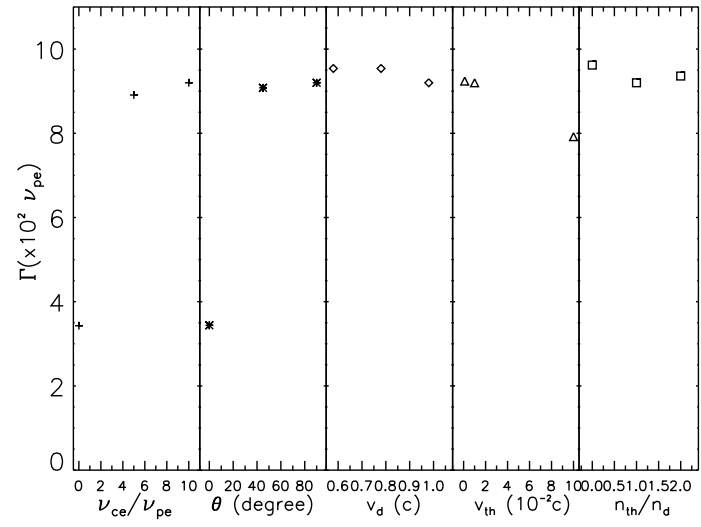}
\caption{The maximum growth rate of the field energy in the relativistic beam-driven instability. }
\label{fig_rgr} 
\end{figure}

The history of the three kind of energies, the polarization of the induced EM field energy and their growth 
rates are illustrated in Fig.\,\ref{fig_renergy}, \ref{fig_rpolarization} and \ref{fig_rgr}. It is seen 
that the background electrons in the simulations have a negligible influence on the induced 
EM field energy and the linear polarization while the thermal energy level of the background electrons 
can suppress the growth of the EM waves. 

From the left panels in Fig.\,\ref{fig_renergy}, we find that the direction and values of the magnetic 
field and the drift velocity can significantly affect the induced EM field. 
The existence of the magnetic field is necessary in order 
to obtain the fast growth of the EM field energy and the energy conversion efficiency seems to have a 
non-linear relation with the magnetic field strength. 

The right panels in Fig.\,\ref{fig_renergy} illustrates the drift energy and the thermal energy of the electrons. 
If the magnetic field strength is sufficiently high to constrain the motion (or spatial position) of the injected electrons, 
the drift energy can be rapidly converted to the thermal energy in a time scale which is similar to the non-relativistic 
case. If the magnetic field is not very strong, we see that it is possible for the system to still retain some residual drift 
energy when the electrons escape from the local region.

The influence of the parameters on the polarization in the relativistic case (Fig.\,\ref{fig_rpolarization}) is 
different from the non-relativistic case. 
The strength of the magnetic field in the present simulations affects the degree of the linear polarization. 
The existence of a strong magnetic field is essential to generate the linear polarization. The non-perpendicular angle 
between the magnetic field and the drift velocity plays an important role in the generation of the linear polarization. In fact, 
at high frequency, linear polarization becomes dominant in the relativistic case. The drift velocity, the thermal level of 
the background electrons and the relative number of drift electrons and background electrons do not affect the polarization 
significantly in the present simulations. 
As shown in Fig.\,\ref{fig_rgr}, the growth rate of the EM field rapidly increases with the magnetic field strength and injection 
angle. However, it seems that in the relativistic case the increase of the drift velocity slightly decreases the growth of the 
EM field perhaps because of a strong interaction between the induced EM field and the high energy electron beam. The thermal 
level of the background electrons can suppress the growth of the EM field, and there is an ambiguous relation between the 
number of background electrons and the growth rate of the EM field in the present simulations. 

\begin{figure*}
\centering
\includegraphics[width=11cm,clip,angle=0]{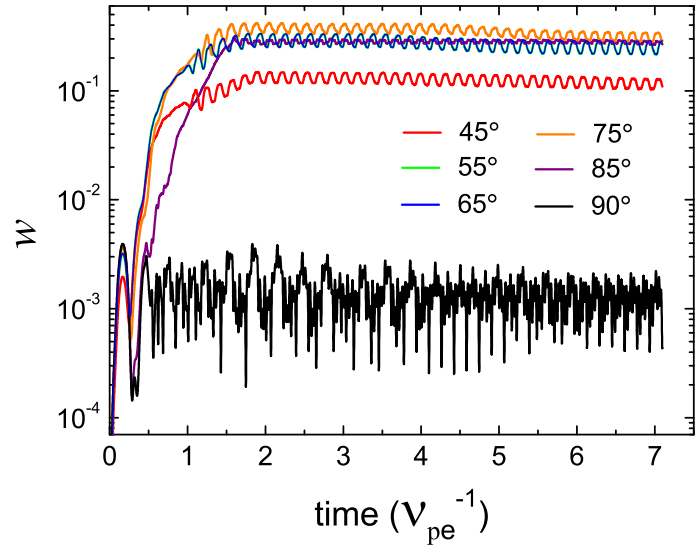}
\caption{The energy history for the standard model with different angle between 
the external magnetic field and the injected velocity direction in the relativistic beam-driven instability, 
i.e. 45$^{\circ}$, 55$^{\circ}$, 65$^{\circ}$, 75$^{\circ}$, 85$^{\circ}$, 90$^{\circ}$. }
\label{fig_srangle} 
\end{figure*}

The influence of the direction between the magnetic field 
and the drift velocity is very interesting since the maximum energy conversion efficiency occurs in the 
non-perpendicular injection which differs from the non-relativistic case. In order to determine the direction 
where we can obtain the maximum energy conversion efficiency, we have done some extra computations by varying 
the direction of the magnetic field. Figure\,\ref{fig_srangle} illustrates the results in which we see 
that the maximum energy conversion efficiency occurs at $\sim75^{\circ}$. 

An interesting phenomenon in our simulations is that the velocity distribution of the beam electrons may evolve from an 
initially drifted Gaussian (or Gaussian-like) distribution to ring distribution (or incomplete-ring or spiral-ring or horse-shoe, e.g. 
see the standard models in both the relativistic case and non-relativistic case.) (A movie is available for the velocity evolution 
of the electron beam in the relativistic case at URL: http://www.arm.ac.uk/$\sim$syu/2decm/relativistic\_beam/\\
relativistic\_beam\_velocity/) 
or spherical-shell distribution (or incomplete shell, e.g. if we change the angle between the magnetic field and the beam electron from 
perpendicular, e.g. 90$^{\circ}$, to non-perpendicular, e.g. 45$^{\circ}$ or 75$^{\circ}$.) This may imply that the ring and shell 
(or ring-like and shell-like) velocity distribution would have a common origin $-$ beam distribution. The influence of the angle on the 
evolution of the velocity of electrons is mainly caused by the external magnetic field and the self-induced EM fields. The spatial 
current induced by the motion of the charged particles plays an important role in the process, which initially causes the variation 
of a spatial magnetic field and time-dependent electric field, i.e. Eq.\,\ref{eq_ampere}. This electric field will accelerate or decelerate 
the electrons to deform the distribution values of the electron velocity. 

\subsection{Spectrum}
\label{sec_rspectrum}

\begin{figure*}
\centering
\includegraphics[width=11cm,clip,angle=90]{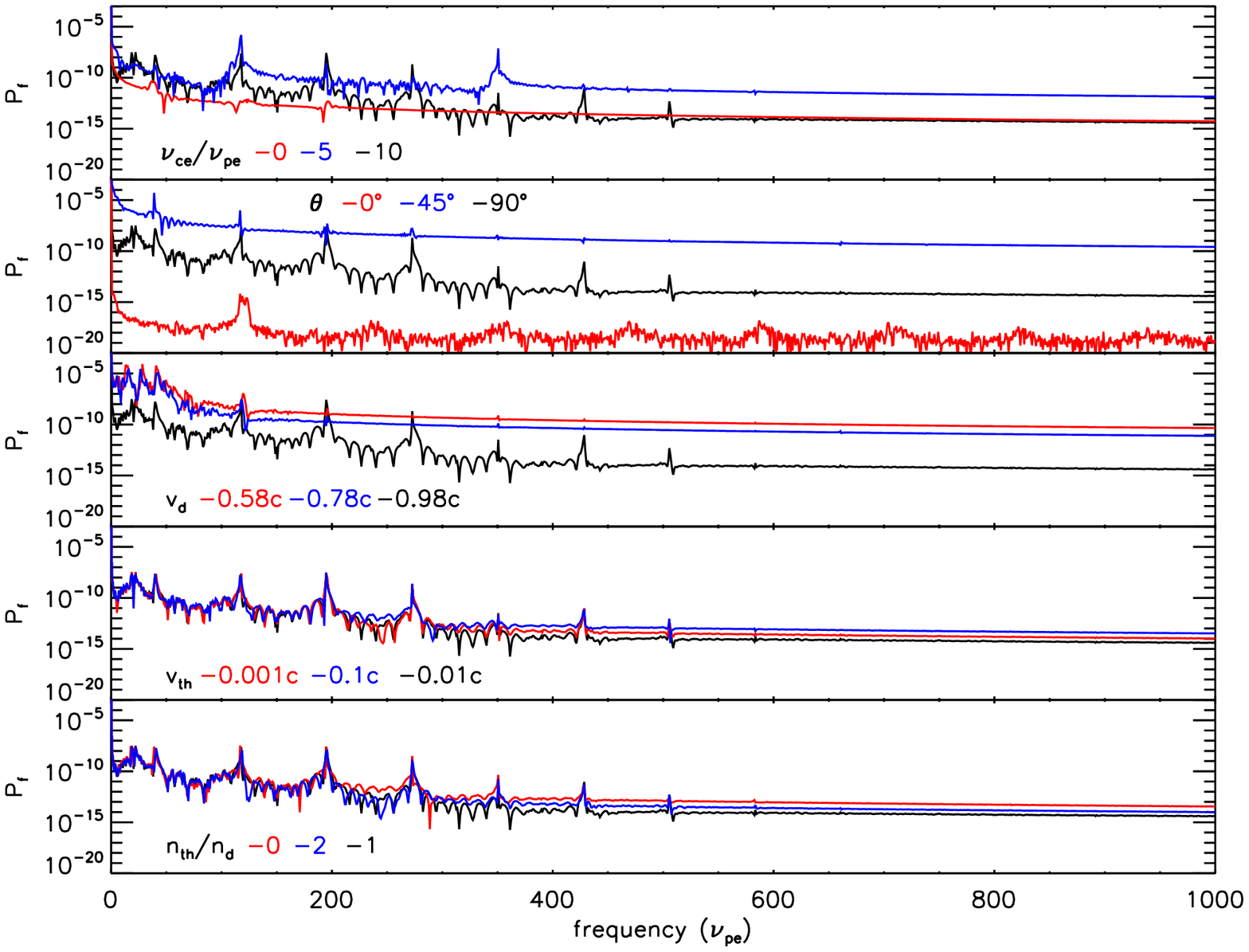}
\caption{Same as Fig.\,\ref{fig_gspectrum} but for the relativistic beam-driven instability. }
\label{fig_rspectrum} 
\end{figure*}

The SED of the radiation in the relativistic electron beam apparently differs from that in the non-relativistic 
case. Figure\,\ref{fig_rspectrum} illustrates the SEDs in the relativistic case. In the standard model, we clearly 
see that the energy may distribute not only in the range of 10 to 100$\nu_{\rm pe}$, but 
extend up to a frequency of $\sim500\nu_{\rm pe}$ and some negligible signal at an even higher frequency harmonic. 

For a single relativistic electron, the peak value of its synchrotron radiation may be at 
$\nu_{\rm peak}\approx\gamma^{2}\nu_{\rm ce}\sin\theta$ \citep{Rybicki79}. For a many electron system, the 
frequencies where we can obtain emission lines is strongly associated with the distribution of the Lorentz 
factor $\gamma$. Figure\,\ref{fig_edis} illustrates the distribution of $\gamma$ in different parameters. 
In the standard model, we find that when the instability reaches saturation the distribution of 
$\gamma$ expands only a little in the high energy part. However, when varying the angle $\theta$ from 
$90^{\circ}$ to $75^{\circ}$, we see a distinct high energy electron tail around $\gamma=9.5$ which 
corresponds to the kinetic energy $E_{i\rm ek}=(\gamma-1)m_{\rm e}c^{2}\approx4.35$ MeV. These high-energy 
electrons may contribute to possible X-ray emission from UCDs via thermal or non-thermal bremsstrahlung 
or even inverse Compton scattering. 

\section{Comparison with observations and other studies}
\label{sec_comparison}

It is possible to approximately compare our simulation results with the observed radio spectrum from UCDs using 
some reference values of $\nu_{\rm pe}$ in Table\,\ref{tab_plasma}. For example, if we assume that the observed 
radio emission from TVLM 513-46546 at 8.5 GHz corresponds to the simulated peak frequency at $\sim12\nu_{\rm pe}$ 
in the case of $\nu_{\rm ce}/\nu_{\rm pe}=10$ (the strongest in the simulation) in the non-relativistic beam-driven 
instability, then $\nu_{\rm pe}=0.708$ GHz and $\nu_{\rm ce}=7.08$ GHz. Thus the corresponding local plasma density 
is $\sim6.2\times10^{9}{\rm cm}^{-3}$ and the strength of the magnetic field is $\sim2529$ Gauss. We may find some 
weaker signal at a high frequency harmonic, such as $\sim30\nu_{\rm pe}=21.24$ GHz, $\sim42\nu_{\rm pe}=29.74$ GHz, 
$\sim50\nu_{\rm pe}=35.4$ GHz. 

In the relativistic case, the first significant signal appears at $\sim20\nu_{\rm pe}$ when $\nu_{\rm ce}/\nu_{\rm pe}=10$, 
which might correspond to the observed radio signal at 8.5 GHz. This gives $\nu_{\rm pe}=0.425$ GHz and 
$\nu_{\rm ce}=4.25$ GHz. Thus the corresponding local plasma density is $\sim2.2\times10^{9}{\rm cm}^{-3}$ and the 
strength of the magnetic field is $\sim1518$ Gauss. We may find some weaker signal at a higher frequency harmonic, such as 
$\sim40\nu_{\rm pe}=17$ GHz, $\sim118\nu_{\rm pe}=50.15$ GHz, $\sim195\nu_{\rm pe}=82.875$ GHz, etc. This means the spectrum 
may extend to the extreme high frequency band or even far infrared in the extremely relativistic case.

If we choose $\nu_{\rm ce}/\nu_{\rm pe}=5$ in the relativistic case, and assume that the first significant 
signal appearing at $\sim110\nu_{\rm pe}$ corresponds to the observed radio signal at 8.5 GHz, it gives $\nu_{\rm pe}=0.0773$ GHz and 
$\nu_{\rm ce}=0.386$ GHz. Thus the corresponding local plasma density is $\sim7.4\times10^{7}{\rm cm}^{-3}$ and the 
strength of the magnetic field is $\sim138$ Gauss.

The above estimations of the properties of the emission region and the spectrum distributions in the present simulations 
are consistent with the analytical results in \citet{Dulk85} and \citet{Guedel02}. 

We do not draw conclusions on the values of plasma parameters in this paper because firstly many parameters are free 
in our simulations. Especially two of the key parameters, the local electron density and magnetic field, are ambiguous. 
Secondly the present simulations are confined in a local micro-region with no consideration of absorption and 
emission by other regions. The absorption and re-emission by the plasma in other regions needs to be further investigated. 
The radio spectrum can be affected by the large-scale structure of a magnetic field \citep{Kuznetsov11}. A combination of 
observations at other wavelengths will help determine the configuration of the magnetic field and the plasma environment, 
and thus entirely understand the magnetosphere and atmosphere of ultracool dwarfs.

The simulation results of the radio pulses from one of the radio active UCDs in \citet{Yu11} show that the loss-cone-driven ECM 
in \citet{Aschwanden90a} can result in the release of $\sim$0.5\% of the kinetic energy, which therefore can generate a strong radio 
flux of up to a few mJy. This depends on other parameters, for instance the local plasma density. By comparison, the non-relativistic 
beam-driven ECM in the present model can release $\sim$2\% of the kinetic energy when the process reaches saturation, which is about 
4 times higher than the results in the model of \citet{Aschwanden90a}. 

Our simulation results are also comparable with the studies for auroral kilometric radiation regions. 
\citet{Pritchett02} investigated the auroral kilometric radiation source cavity using two-dimensional 
particle-in-cell simulations. They found the energy conversion efficiency is $\sim1-5$\%, depending on 
the velocity of the hot electrons ($\sim1-10$ keV). The time scale of the radiation bursts is $\sim$0.5 ms, 
which is probably associated with the local plasma density ($\lesssim$1 cm$^{-3}$, \citet{Strangeway98,Perraut90,Calvert81}) 
and the weak magnetic field. The energy conversion efficiency in our results are in line with those in their results 
in the non-relativistic cases. The spectra obtained in the present simulations are also in good agreement with 
the previous results \citep{Melrose82,Aschwanden90a,Pritchett86,Pritchett84} - the waves are produced mainly near the 
electron cyclotron frequency. 

However, certain differences exist between our simulation results and others. 
Firstly, in our simulations, we can see the different oscillation modes from the energy 
time history which reflect the interaction between the particles and the induced electric 
field. Secondly, we find that the energy conversion efficiency can be influenced significantly 
by the drift velocity of the particles and the angle between the drift velocity and the magnetic field, 
which is perhaps caused by the self-induced electromagnetic damping. 
The initial conditions and the technique to solve the Maxwell$'$s equations in our simulations and 
others are also different. We considered a spatially localized electron population with a beam or 
beam-like velocity distribution. In contrast, previous works considered ring-like, horseshoe, 
and DGH distributions \citep{Pritchett02,Pritchett86,Pritchett84} and a larger space structure for the 
magnetic field. 
The time advancement of Maxwell$'$s equations was performed in Fourier space in \citet{Pritchett02}. 
Instead, we used a staggered leap-frog method to solve the equations in the time domain with the positions and velocities 
of the electrons generated by a Monte Carlo method. 

Recent computer simulations and experimental laboratory work by \citet{Cairns11} and \citet{Vorgul11} show that in the 
non-relativistic (or weakly relativistic) case the energy conversion efficiency is 1$-$2\%, which is consistent with our simulations. 
However, this energy conversion efficiency varies significantly in our relativistic case, depending on the parameters, e.g. the angle, 
the drift velocity, the strength of external magnetic field. It may be mainly caused by the effect of synchrotron radiation. 

\section{Summary and conclusions}
\label{sec_conclusion}

\begin{figure*}
\centering
\includegraphics[width=17cm,clip]{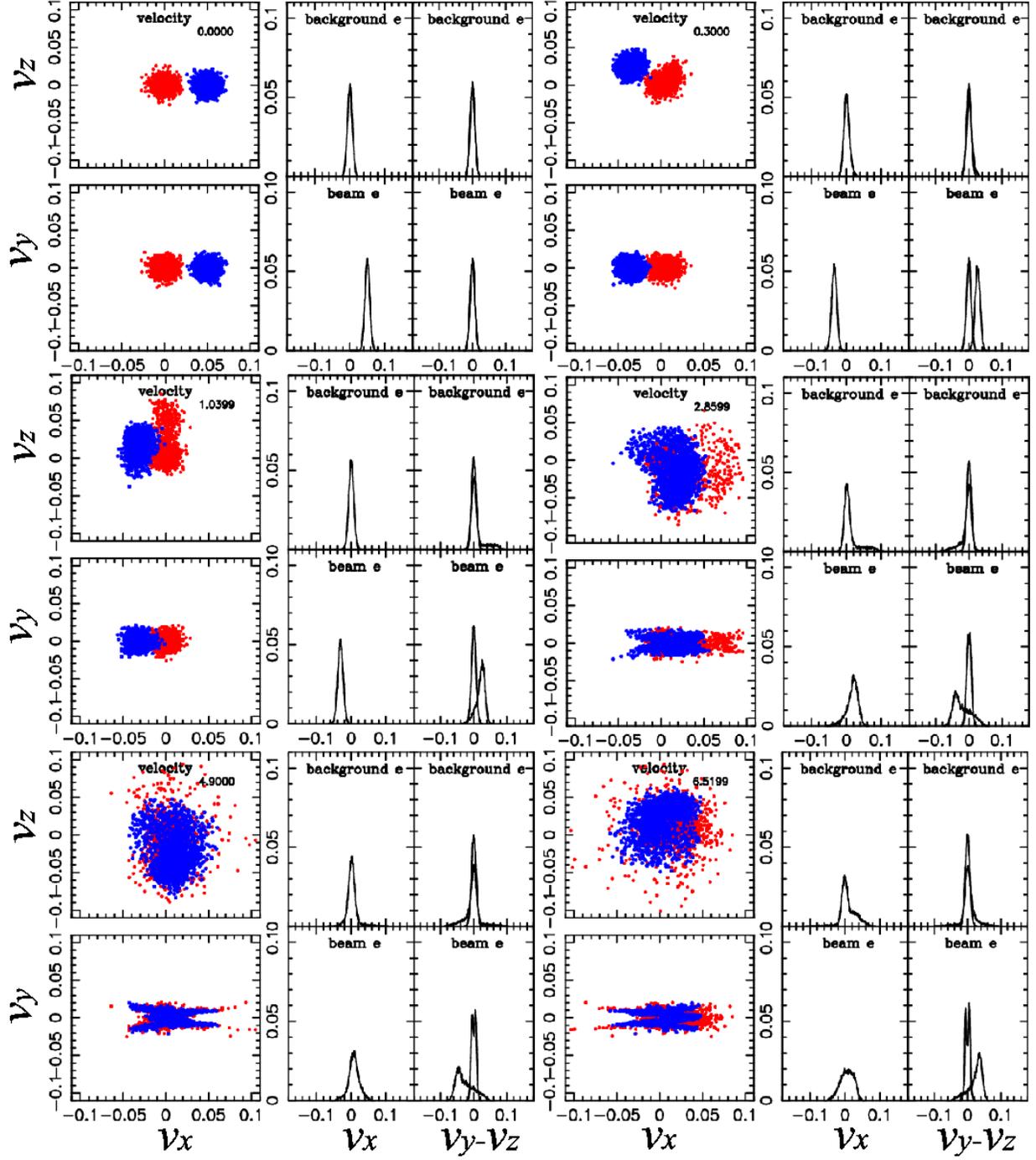}
\caption{The history of the velocity distribution of the background electrons (red points) and beam electrons (blue points). 
Different panels from top-left to bottom-right represent the snapshots for time $t=0.00$, 0.30, 1.04, 2.86, 4.90 and 6.52 $\nu_{\rm pe}^{-1}$. }
\label{fig_velocity1} 
\end{figure*}

\begin{figure}
\centering
\includegraphics[width=12cm,clip,bb=20 15 490 245,angle=0]{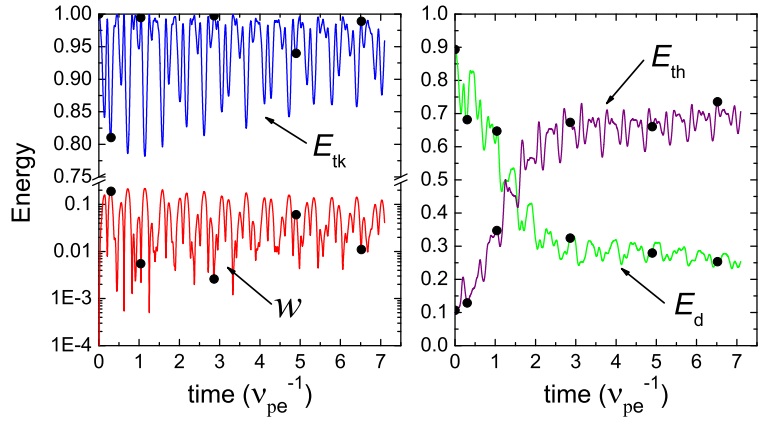}
\caption{The energy history of the standard model. Blue line in the left panel denotes the total kinetic energy, 
and red line stands for the field energy. In the right panel, purple line is for the total thermal energy of the system, 
and the green line is for the drift energy of the injected electrons. Black points denote the time as in 
Fig.\,\ref{fig_space1} and \ref{fig_velocity1}.}
\label{fig_senergy} 
\end{figure}

\begin{figure}
\centering
\includegraphics[width=12cm,clip,bb=16 10 380 230,angle=0]{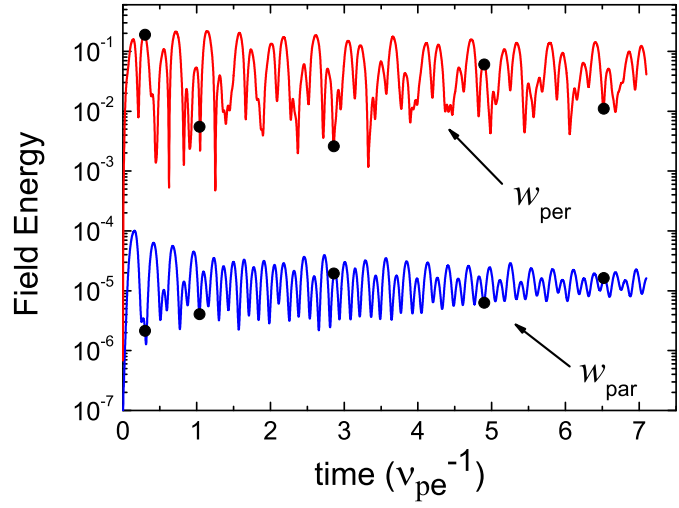}
\caption{The history of the field energy along the direction of perpendicular (red line) and 
parallel (blue line) to the external magnetic field. Black points denote the time as in 
Fig.\,\ref{fig_space1} and \ref{fig_velocity1}.}
\label{fig_spolarization} 
\end{figure}

\begin{figure*}
\centering
\includegraphics[width=11cm,clip,angle=0]{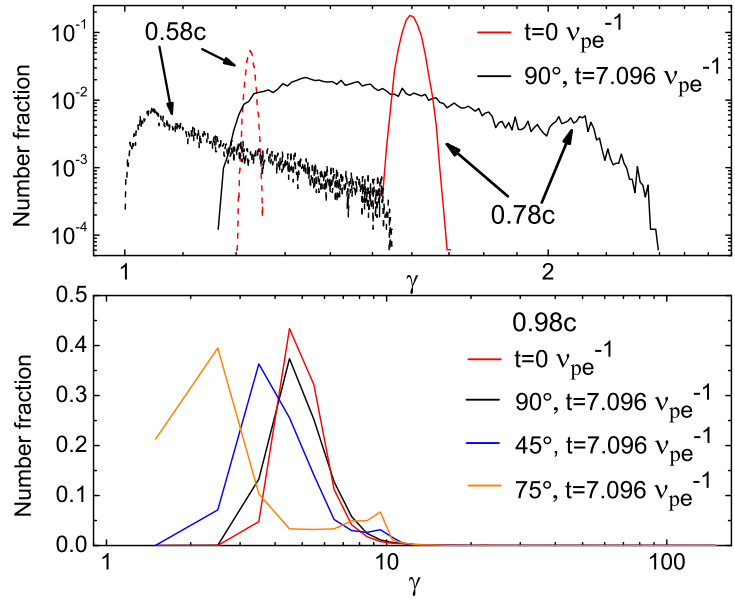}
\caption{The energy distribution of the electrons in the relativistic beam-driven instability. Top panel is for the 
initially mildly relativistic electrons with drift velocity 0.58$c$ and 0.78$c$, and incident angle 90$^{\circ}$; 
the bottom panel is for initially mildly relativistic electrons with fixed drift velocity 0.98$c$ and different 
incident angle, i.e. 45$^{\circ}$, 75$^{\circ}$, 90$^{\circ}$. The number of the electrons in each bin is normalized 
by the total number of injected electrons. The bin size is 0.001 for $0.58c$, 0.01 for $0.78c$, and 1 for $0.98c$.}
\label{fig_edis} 
\end{figure*}

In this paper, we present the numerical simulations for electron-beam-driven and loss-cone-driven 
ECM with different plasma parameters and different magnetic field strengths. We find that the 
beam-driven ECM can be an effective mechanism to release EM waves and heat the surrounding plasmas. 
From the diffusion process of the electrons in velocity space and the energy distribution 
(see Fig.\,\ref{fig_velocity1} and Fig.\,\ref{fig_edis}), a high-energy tail of the electrons 
may be rapidly developed along the direction near perpendicular to the magnetic 
field, which can eventually evolve to moderately or strongly relativistic electrons depending 
on the initial energy of the electron current, and contribute towards gyrosynchrotron or synchrotron 
radiation. This may lead to the appearance of a radio continuum and the deformation of the SED. 
Also, these high-energy electrons may be important to generate X-ray emission. 

The computation of the degree of polarization indicates that the thermal level of the electrons can 
significantly affect the degree of the circular or linear components of the observed radio waves. 
In the case of the beam-driven ECM, the angle between the direction of the magnetic field and the 
injection direction of the injected electrons is another crucial factor to affect the degree 
of circular polarization in the radio waves. 

The SEDs of the radio waves depends weakly on the form of the velocity distribution of the 
electrons in the present simulations. The existence of the external magnetic field and the 
angle between the direction of the magnetic field and the moving direction of the electron 
current can significantly affect the SED. Certain frequency bands, e.g.  
10 to 70$\nu_{\rm pe}$ in the non-relativistic case and 
10 to 600$\nu_{\rm pe}$ in the relativistic case may appear, which increases 
the difficulty of finding the fundamental cyclotron frequency in the observed radio frequencies. 
It is however possible that magnetic field inhomogeneities may smooth out some of 
these bands thus producing a continuous spectrum. 
In order to determine the plasma frequency and the cyclotron frequency, wide frequency bands 
should be covered by future radio observations. 

The present study is limited in that only two-dimensional electromagnetic simulations are performed with 
no consideration of the change of the magnetic field configuration and the influence of gravity. 
We will continue to develop the simulations to match the plasma environment and the magnetic 
topology on UCDs in order to understand their radio emission.

\acknowledgments
Research at the Armagh Observatory is grant-aided by the N. Ireland Department 
of Culture, Arts and Leisure. We also thank the Leverhulme Trust for their 
support of this project. AA gratefully acknowledges the support of the Bulgarian 
National Science Fund (grant no. DDVU02/40 - 2010). ALM thanks STFC for support 
through a Rolling Grant. SY thanks the 9th international space simulation school 
for the tutorial courses and Prof. Y. Omura for discussions. SY also thanks 
Dr. J. McDonald in NUI, Galway for discussions. We thank the referee for his/her helpful 
and constructive comments. 

\bibliographystyle{harvard.sty}
\bibliography{ucda}
\clearpage

\clearpage

\clearpage

\clearpage

\clearpage

\clearpage

\clearpage

\clearpage

\clearpage

\clearpage

\clearpage

\clearpage

\clearpage

\clearpage

\clearpage

\clearpage

\clearpage

\clearpage

\clearpage

\clearpage

\begin{table}
\begin{minipage}[t]{\columnwidth}
\caption{Reference values of electron densities and related parameters.}
\label{tab_plasma}
\begin{center}
\begin{tabular}{lccccccccc}
\hline
       n$_{\rm e}$              & $\nu_{\rm pe}$                         &    $\lambda _{\rm D}$      & $v_{\rm th}/$Temperature                 \\
     (cm$^{-3}$)                &  (s$^{-1}$)                            &  (cm)                      &        (cm s$^{-1}/$K)                   \\
     10$^{12}$                  & 8.98$\times$10$^{9}$                   & 0.005                      &  3$\times$10$^{8}/$3$\times$10$^{5}$      \\
     10$^{10}$                  & 8.98$\times$10$^{8}$                   & 0.05                       &  3$\times$10$^{8}/$3$\times$10$^{5}$      \\
     10$^{8}$                   & 8.98$\times$10$^{7}$                   & 0.5                        & 3$\times$10$^{8}/$3$\times$10$^{5}$      \\
     10$^{6}$                   & 8.98$\times$10$^{6}$                   & 5                          &  3$\times$10$^{8}/$3$\times$10$^{5}$     \\
     10$^{4}$                   & 8.98$\times$10$^{5}$                   & 50                         & 3$\times$10$^{8}/$3$\times$10$^{5}$      \\
     10$^{2}$                   & 8.98$\times$10$^{4}$                   & 500                        & 3$\times$10$^{8}/$3$\times$10$^{5}$       \\
\hline   
\end{tabular}
\end{center}
\end{minipage}
\end{table}

\clearpage

\begin{table}
\begin{minipage}[t]{\columnwidth}
\caption{Various parameters in the simulations (\S\,\ref{sec_resultsI}) and their values. 
Note that part of the values may vary in \S\,\ref{sec_resultsII} and \S\,\ref{sec_relativistic}.}
\label{tab_parameters}
\begin{center}
\begin{tabular}{lccccccccc}
\hline
Parameters                                       & Symbols                          & Standard         & Optional              \\
                                                 &                                  &  values          &  values               \\
Plasma frequency                                 & $\nu_{\rm pe}$                   & 1                & -                    \\
Cyclotron frequency                              & $\nu_{\rm ce}$                   & 10               & 0, 5                 \\
Angle: $\textbf{B}_{0}$ and $\textbf{v}_{\rm d}$     & $\theta$                     & 90$^{\circ}$     & 0, 45$^{\circ}$      \\
Thermal velocity \\
of background $e$                                & $v_{\rm th}$            & 0.01$c$          & 0.005$c$, 0.05$c$             \\
Thermal velocity \\
of drift $e$                                     & $v'_{\rm th}$           & 0.01$c$          & 0.005$c$, 0.05$c$             \\
Drift velocity                                   & $\textbf{v}_{\rm d}$             & 0.05$c$          & 0.0$c$, 0.1$c$        \\
Number of superparticles \\
of background $e$                                & $n_{\rm th}$                     & 16384            & 0, 32768               \\
Number of superparticles \\
of drift $e$                                     & $n_{\rm d}$                      & 16384            & -                       \\
\hline   
\end{tabular}
\end{center}
\end{minipage}
\end{table}

\end{document}